\documentclass[aps,prl,twocolumn,groupedaddress,floatfix,showpacs]{revtex4-1}

\usepackage{graphicx}
\usepackage{amsmath}
\usepackage{amssymb}
\usepackage{natbib}

\providecommand{\abs}[1]{\lvert#1\rvert}
\newcommand{\vect}[1]{\mathbf{#1}}

\begin{document}


\title{Predictability and suppression of extreme events in a chaotic system}


\author{Hugo L. D. de S. Cavalcante}
\email[To whom correspondence should be addressed.]{ hugo.cavalcante@pq.cnpq.br}
\altaffiliation[Present Address: ]{Departamento de Inform\'atica, Universidade Federal da Para\'{\i}ba, CEP  58051-900 - Jo\~{a}o Pessoa, PB - Brazil.}
\affiliation{Grupo de F\'{\i}sica At\^omica e Lasers - DF, Universidade Federal da Para\'{\i}ba, Caixa Postal 5086 - 58051-900 - Jo\~{a}o Pessoa, PB - Brazil.}
\author{ Marcos Ori\'{a}}
\affiliation{Grupo de F\'{\i}sica At\^omica e Lasers - DF, Universidade Federal da Para\'{\i}ba, Caixa Postal 5086 - 58051-900 - Jo\~{a}o Pessoa, PB - Brazil.}
\author{Didier Sornette}
\affiliation{ETH Zurich, Department of Management, Technology and Economics, Scheuchzerstrasse 7, CH-8092 Zurich, Switzerland.}
\author{Edward Ott}
\affiliation{Institute for Research in Electronics and Applied Physics, University of Maryland, College Park, MD, 20742 USA.}
\author{Daniel J. Gauthier}
\affiliation{Department of Physics, Duke University, Box 90305, Durham, NC, 27708 USA.}


\date{\today}

\begin{abstract}
 f earthquakes, neuroscience, ecology, and even financial economics.
In many complex systems, large events are believed to follow power-law, scale-free probability distributions, so that the extreme, catastrophic events are unpredictable. Here, we study coupled chaotic oscillators that display extreme events. The mechanism responsible for the rare, largest events makes them distinct and their distribution deviates from a power law. Based on this mechanism identification, we show that it is possible to forecast in real time an impending extreme event. Once forecasted, we also show that extreme events can be suppressed by applying tiny perturbations to the system.
\end{abstract}


\maketitle

Extreme events are increasingly attracting the attention of scientists and decision makers because of their impact on society \cite{Nott2006,Comfort2010,Field2012,Barnosky2012}, which is exacerbated by our increasing global interconnectivity.
Examples of extreme events include financial crises, environmental and industrial accidents, epidemics and blackouts \cite{GlobalRisks2013}. 
From a scientific view point, extreme events are interesting because they often reveal underlying, often hidden, organizing principles \cite{Rundle1996,Sornette2002,Albeverio2005}. 
In turn, these organizing principles may enable forecasting and control of extreme events. 

Some progress along these lines has emerged in studies of complex systems composed of many interacting entities. For example, it was found recently that complex systems with two or more stable states may undergo a bifurcation causing a transition between these states that is associated with an extreme event \cite{Scheffer2012,Biggs2009}. Critical slowing down and/or increased variability of measureable system quantities near the bifurcation point open up the possibility of forecasting an impending event, as observed in laboratory-replicated populations of budding yeast \cite{Dai2012}.

An open question is whether other underlying behaviors cause extreme events.  One possible scenario is when the system varies in time and is organized by attracting sets in phase space.  For example, a recent model of financial systems consisting of coupled, stochastically-driven, linear mappings \cite{Krawiecki2002} shows so-called bubbling behavior, where a bubble -- an extreme event -- corresponds to a large temporary excursion of the system state away from a nominal value.  In this example, the event-size distribution follows a power law, having a ``fat'' tail that describes the significant likelihood of extreme events. One main characteristic of such distributions is that they are scale-free, which means that events of arbitrarily large sizes are caused by the same dynamical mechanisms governing the occurrence of small- and intermediate-size events, leading to an impossibility of forecasting \cite{Sornette2009,Taleb2007,Bak1996,Embrechts2011,Knight1921}.

In contrast, the new concept of ``dragon-kings'' (DKs) emphasizes that the most extreme events often do not belong to a scale-free distribution \cite{Sornette2009}. DKs are outliers, which possess distinct formation mechanisms \cite{Sornette2012}. Such specific underlying mechanisms open the possibility that DKs can be forecasted, allowing for suppression and control. 
Here, we show DK-type statistics occurring in an electronic circuit that has an underlying time-varying dynamics identified to belong to a more general class of complex systems.
Moreover, we identify the mechanism leading to the DKs and show that they can be forecasted in real time, and even suppressed by the application of tiny and occasional perturbations.  
The mechanism responsible for DKs in this specific system is \emph{attractor bubbling}. As explained below, we argue that attractor bubbling is a generic behavior appearing in networks of coupled oscillators, and that DKs and extreme events are likely in these extended systems.

The large class of spatially extended coupled oscillator networks covers the physics of earthquakes \cite{Schmittbuhl1993}, biological systems such as the collective phase synchronization in brain activity \cite{Gong2007}, and even of financial systems made of interacting investors with threshold decisions and herding tendencies \cite{Takayasu1992}. Many coupled-oscillator system models exhibiting chaos have 
invariant manifolds --- subspaces of the entire phase space on which the system trajectory can reside.  
 o another attractor, eventually followed by reinjection to the dominating attractor in many situations. In sum, attractor bubbling associated with riddled basins of attraction is a generic mechanism for DKs, which we conjecture apply to a large class of spatially extended deterministic and stochastic nonlinear systems.	
 Such manifolds commonly occur in models where identical chaotic systems are coupled and synchronize. Furthermore, when invariant manifolds contain chaotic orbits, they can lead to attractor bubbling (as well as riddled basins and on-off intermittency)
 \cite{Ott2002, Sommerer1993, Mosekilde2002, Ashwin1994}. Attractor bubbling 
is a situation where the system trajectory irregularly and briefly leaves the vicinity of an invariant manifold containing a chaotic attractor as a result of an occasional noise-induced jump into a region where orbits are locally repelled from the invariant manifold.
The system state then follows an orbit that moves away from the invariant manifold, but  eventually returns to the attractor.
These excursions of the system state to phase space regions far from the invariant manifold are our extreme events.

To highlight the connection between  attractor bubbling and DKs we study two nearly identical unidirectionally-coupled chaotic electronic circuits in a master (mnemonic $M$) and slave ($S$) configuration. The state of each circuit is described by a three-dimensional (3D) vector whose components are related to the two voltages and the current of each circuit (see Fig.\ S1 \footnote{See Supplemental Material at [URL will be inserted by publisher] for additional figures and details.}).  The temporal evolution of the state vectors is governed by the differential equations  
\begin{align}
\dot{\vect{x}}_M & = \vect{F}\left[ \vect{x}_M \right] , \label{eq:master_flow}\\
\dot{\vect{x}}_S & = \vect{F}\left[ \vect{x}_S \right] 
	+c\vect{K}\left( \vect{x}_M - \vect{x}_s
	\right), \label{eq:slave_flow}
\end{align}
where the dot over a variable means differentiation with respect to time,  $\vect{F}\left[\vect{x}\right]$ is the flow for each subsystem, $c$ controls the interaction strength between the subsystems, and $\vect{K}$ is the coupling matrix.  In general, the coupled system resides in a 6D phase space spanned by $\left(\vect{x}_M,\vect{x}_S\right)$.  However, for appropriate values of $c$ and $\vect{K}$, the coupled oscillators synchronize their behavior \cite{Sommerer1993}, which corresponds to $\vect{x}_M=\vect{x}_S$. Hence, the coupled-system trajectory resides in a restricted 3D subspace (on an invariant manifold). In this case, it is insightful to introduce new 3D state vectors that describe the behavior on the invariant manifold $\vect{x}_{\parallel} = \left(\vect{x}_M+\vect{x}_S\right)/2 $ and transverse to the manifold $\vect{x}_{\perp} = \left(\vect{x}_M-\vect{x}_S\right)/2 $. Synchronization corresponds to $\vect{x}_{\perp} = 0$  and  $\vect{x}_{\parallel} = \vect{x}_M
 $, and the basin of attraction associated with the synchronized state is riddled.

Here we study a pair of electronic circuits for which the equations (\ref{eq:master_flow}) and (\ref{eq:slave_flow}) take the form 
\begin{align}
\dot{V}_{1j} & = \frac{V_{1j}}{R_1} -g\left[ V_{1j}-V_{2j}\right] , \label{eq:V1dot}\\
\dot{V}_{2j} & = g\left[ V_{1j}-V_{2j}\right] -I_j
	+\delta_{S,j} c\left( V_{2M} - V_{2S}
	\right), \label{eq:V2dot}\\
\dot{I}_j & = V_{2j} -R_4 I_j,  \label{eq:Idot} 
\end{align}
for $j = M, S$, respectively, $\delta_{S,j}$  is the Kronecker delta ($1$ if $j=S$ and $0$ if $j=M$), and 
\begin{equation}
g\left[V\right] = \frac{V}{R_2} +I_r \left( \exp(\alpha_f V)-\exp(-\alpha_r V) \right).
\end{equation}
The values of the parameters and other details are given in \cite{Note1,Gauthier1996}.  Equations (\ref{eq:V1dot}-\ref{eq:Idot}) correspond to Eqs.\ (\ref{eq:master_flow}-\ref{eq:slave_flow}) with $\vect{x}_j = (V_{1j}, V_{2j}, I_j)^T$, where $\vect{x}^T$ denotes the transpose of vector $\vect{x}$, and the coupling matrix $\vect{K}$ is such that the matrix entry $K_{m,n}$  is $1$ for $m=n=2$ and $0$ otherwise.

\begin{figure}[tbhp]
\includegraphics[scale=0.85]{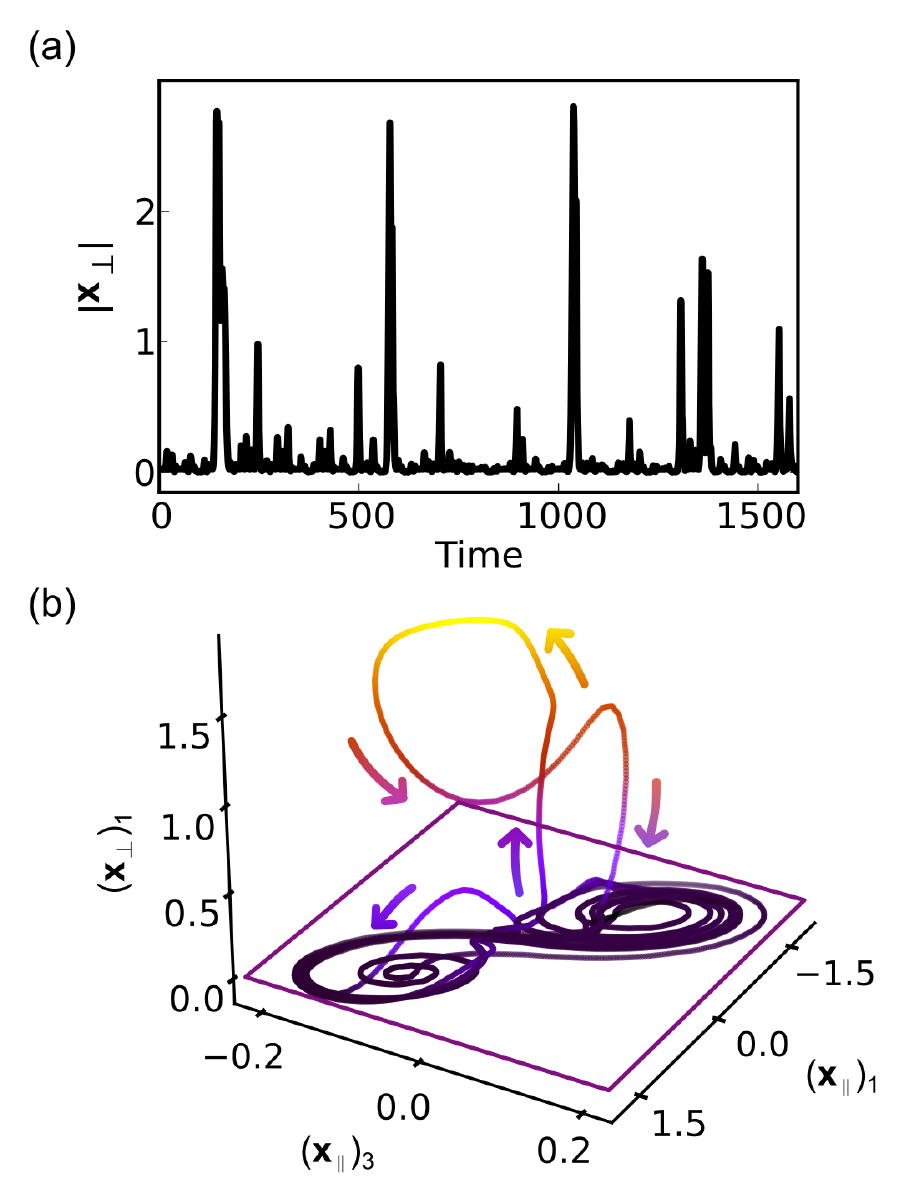}
\caption{(color online). Experimental observation of attractor bubbling in coupled chaotic oscillators. (a) Typically observed temporal evolution of $\abs{\vect{x}_{\perp}}$ for $c$ = 4.4. 
Time is measured in dimensionless units corresponding to the characteristic time in the circuit (see Supplemental Material). (b) Illustration of the system trajectory in the vicinity of a bubbling event. The 6D phase space is projected onto a 3D space, where the subscript on the axis label indicates the $i^\mathrm{th}$ component of the corresponding vector. The arrows indicate the direction of the flow and the colors indicate the height on the $\left(\vect{x}_{\perp}\right)_1$  direction. \label{fig:bubbles}}
\end{figure}
As discussed above, attractor bubbling occurs when noise is present (e.g., thermal noise in the electronic components), when there is a slight parameter mismatch between the oscillators (the flows of each circuit are slightly different), or when both effects are present, which is the most likely situation in an experiment. Bubbling is indicated by long excursions of high-quality synchronization ($\vect{x}_{\perp}$  close to the noise level) interspersed by brief desynchronization events where $\vect{x}_{\parallel}$  takes on a large value --- an extreme event --- as shown in Fig.\ \ref{fig:bubbles}(a).  We illustrate the trajectory of a typical bubbling event in Fig.\ \ref{fig:bubbles}(b), which is a projection of the 6D phase space onto a 3D space containing components of the invariant manifold and of the transverse manifold.  It is seen that the trajectory remains for most of the time on the invariant manifold $\vect{x}_{\parallel}$, but undergoes a large excursion away fro
 m the invariant manifold during the bubbling event.  Due to the nonlinear folding of the flow, the trajectory is reinjected to the invariant manifold after the bubble.

\begin{figure}[tbh]
\includegraphics[scale=0.95]{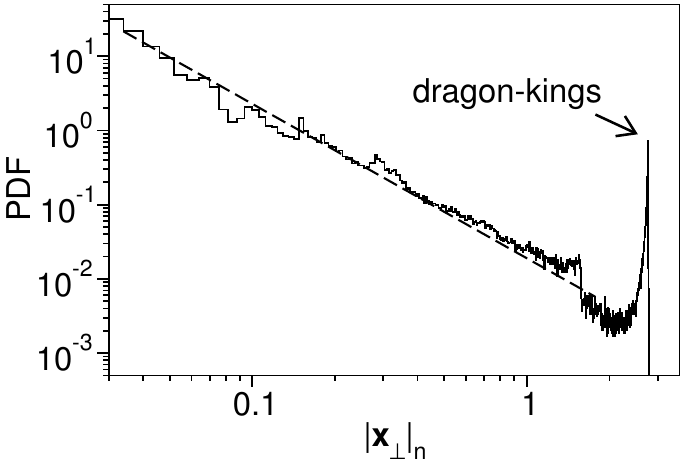}
\caption{Appearance of dragon-kings. Bubble event-size probability density function (PDF) for $c$ = 4.4.
The dashed line is a fit to a power law. 
\label{fig:pdf}}
\end{figure}

To reveal the existence of DKs, we collect a long time series of values of $\abs{\vect{x}_{\perp}}$, use a peak-detecting algorithm to identify the bubbling events, and create a probability density function (PDF) for the event-sizes $\abs{\vect{x}_{\perp}}_n$, defined as the largest peak-value of $\abs{\vect{x}_{\perp}}$ within a burst. 
The length of the time series is large enough that the observed PDFs have reached statistical convergence and are stationary, in the sense that their shape does not change appreciably with the addition of new samples.
The resulting distribution is shown in Fig.\ \ref{fig:pdf}, where 
the event-sizes follow approximately a power-law distribution (straight line in log-log scale) for small to moderately large sizes ($0.04 < \abs{\vect{x}_{\perp}}_n <1.8$) with exponent $-2.0\pm 0.1$.  
We apply a Kolmogorov-Smirnov (K-S) statistical hypothesis test to check that the distribution of event sizes follows a power law in this interval. 
The hypothesis of a truncated power-law is rejected for the raw data because there are small but statistically significant deviations from a straight line decorating the distribution. The hypothesis of a truncated power-law is accepted with the same value of the exponent obtained in the fit if we apply a decimation of correlated data by resampling the raw data \cite{Note1}.
This empirical observation is substantiated by a theoretical analysis based on the statistics of the perturbations affecting trajectories near the fixed point at the origin  \cite{Note1}.
The analysis predicts that the exponent is $-2$ to leading order. Moreover, the observed desynchronization events can be rationalized as being associated with the structure of the repeller around the origin, consistent with current theory of attractor bubbling.

A substantial and significant peak in the distribution and subsequent cut-off that deviates from the power law is observed for the extremely large events ($\abs{\vect{x}_{\perp}}_n>2.4$), which we associate with dragon-kings.
Interestingly, the probability mass contained in the large peak associated to the DKs is approximately equal to the integral of the PDF that would result if the power law extended to infinity. This fact suggests that the DKs are events that would belong to a power-law distribution but had their size limited by some saturation mechanism that effectively determines a maximum size for the events in the system. 
The K-S hypothesis test verifies that this large peak in the PDF deviates significantly from a pure power-law, as expected from theory, using either the raw data or the decimated data \cite{Note1}. Hence, the theory developed assuming linearization near the fixed point captures the essence of the bubbling (power law with exponent $-2$ and dragon-king peak of the PDF), and only fails to explain the tiny structures decorating the distribution. 

As discussed some time ago \cite{Ashwin1994,Heagy1995}, a bubbling event is initiated by ``hot spots'' within the chaotic attractor that resides on the invariant manifold. The attractor is composed of a large (likely infinite) number of unstable sets, such as unstable fixed points, unstable periodic orbits, etc. \cite{Ott2002}. Each of these sets has an associated local transverse Lyapunov exponent \cite{Heagy1995}, which describes the tendency of a trajectory to be attracted to or repelled from the invariant manifold when it is in a neighborhood of the set. A system with attractor bubbling necessarily has a distribution of local Lyapunov exponents (see Fig.\ S2 \cite{Note1}), where at least some are repelling (value greater than zero), even though the value of the weighted average is negative (attracting). The repelling sets correspond to the hot spots on the invariant manifold.

For the coupled oscillators studied here, it was found previously that one set in particular --- the unstable, saddle-type fixed point at $\vect{x}_{\parallel}=0$  ---  is exceedingly transversely unstable and is the underlying originator of the largest bubbles \cite{Gauthier1996,Venkataramani1996}. That is, there is a very high likelihood that a bubble will occur whenever $\vect{x}_{\parallel}$ resides in a neighborhood of the origin for some time, and the largest events (the DKs) occur when the residence time is long and the approach is close.  The large bubble event shown in Fig.\ \ref{fig:bubbles}(b) clearly originates near  $\vect{x}_{\parallel}=0$. 
This observation is at the heart of the theoretical approximation for the distribution of event sizes, where we approximate the dynamics of perturbations by linearization of the equations of motion (Eqs.\ (\ref{eq:V1dot}) to (\ref{eq:Idot}), for $j=M,S$) near the fixed point.

\begin{figure}[!thbp]
\includegraphics[scale=0.95]{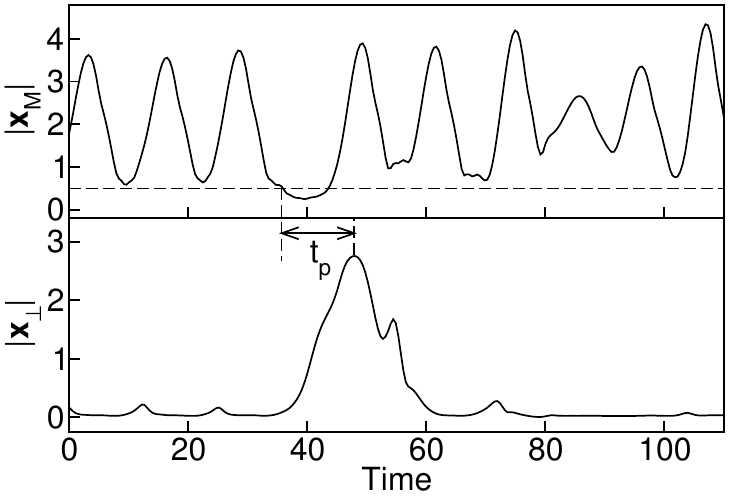}
\caption{Forecasting dragon-kings. Temporal evolution of the trajectory projected onto the invariant manifold ($\abs{\vect{x}_{M}}$) and the transverse space ($\abs{\vect{x}_{\perp}}$) during attractor bubbling ($c$ = 4.4). 
The largest, extreme event, which is part of the dragon-king distribution, is preceded by a long excursion of $\abs{\vect{x}_{M}}$ below a threshold of $\abs{\vect{x}_{M}}_\mathrm{th} = 0.50$ whose value is determined empirically by minimizing the number of false predictions. \label{fig:forecasting}}
\end{figure}

The influence of the fixed point in the dynamics also allows us to predict the occurrence of a large event by real-time observation of  $\vect{x}_{M}$, which is equal to $\vect{x}_{\parallel}$ when the subsystems are synchronized, and finding the times when it approaches the origin. Figure\ \ref{fig:forecasting} shows the temporal evolution of $\abs{\vect{x}_{M}}$  and $\abs{\vect{x}_{\perp}}$, where it is seen that $\abs{\vect{x}_{M}}$ undergoes a sustained drop and remains below an empirically determined threshold value $\abs{\vect{x}_{M}}_\mathrm{th}$  preceding a large bubble (spike in $\abs{\vect{x}_{\perp}}$), where the forecasting time is denoted by $t_p$. A smaller threshold is associated with a larger event size and hence it can be adjusted to isolate the DKs. 
This description and ensuing results are confirmed by numerical integration of Eqs.\ (\ref{eq:V1dot}-\ref{eq:Idot}), which shows excellent agreement with the experimental observations \cite{Note1} and demonstrates that it is possible to forecast DKs in this relatively low-dimensional complex system.

With this scheme to forecast DKs, we design a feedback method to suppress them based on occasional proportional feedback of tiny perturbations to the slave oscillator when $\abs{\vect{x}_{M}} < \abs{\vect{x}_{M}}_\mathrm{th}$  \cite{Newell1994}. In the presence of feedback, the temporal evolution of the slave oscillator (Eq.\ (\ref{eq:slave_flow})) is modified to read 
\begin{align}
\dot{\vect{x}}_S   = & \vect{F}\left[ \vect{x}_S \right]  
	 +c\vect{K}\left( \vect{x}_M - \vect{x}_s \right) \label{eq:DK_feedback} + \\
	& \left[1-\theta\left(\abs{\vect{x}_{M}}-\abs{\vect{x}_{M}}_\mathrm{th}\right) \right]
	   c_{DK}\vect{K}_{DK}\left( \vect{x}_M - \vect{x}_s 
	 \right), \notag
\end{align}
where $\theta$ is the Heaviside step function, and $c_{DK}$ ($\vect{K}_{DK}$) is the feedback strength (coupling matrix) used to suppress DKs. For the purpose of illustration, we assume that it is expensive or not convenient to keep this additional feedback coupling on all the time and thus it is only active for a brief interval when necessary.

\begin{figure}[!htbp]
\includegraphics[scale=0.90]{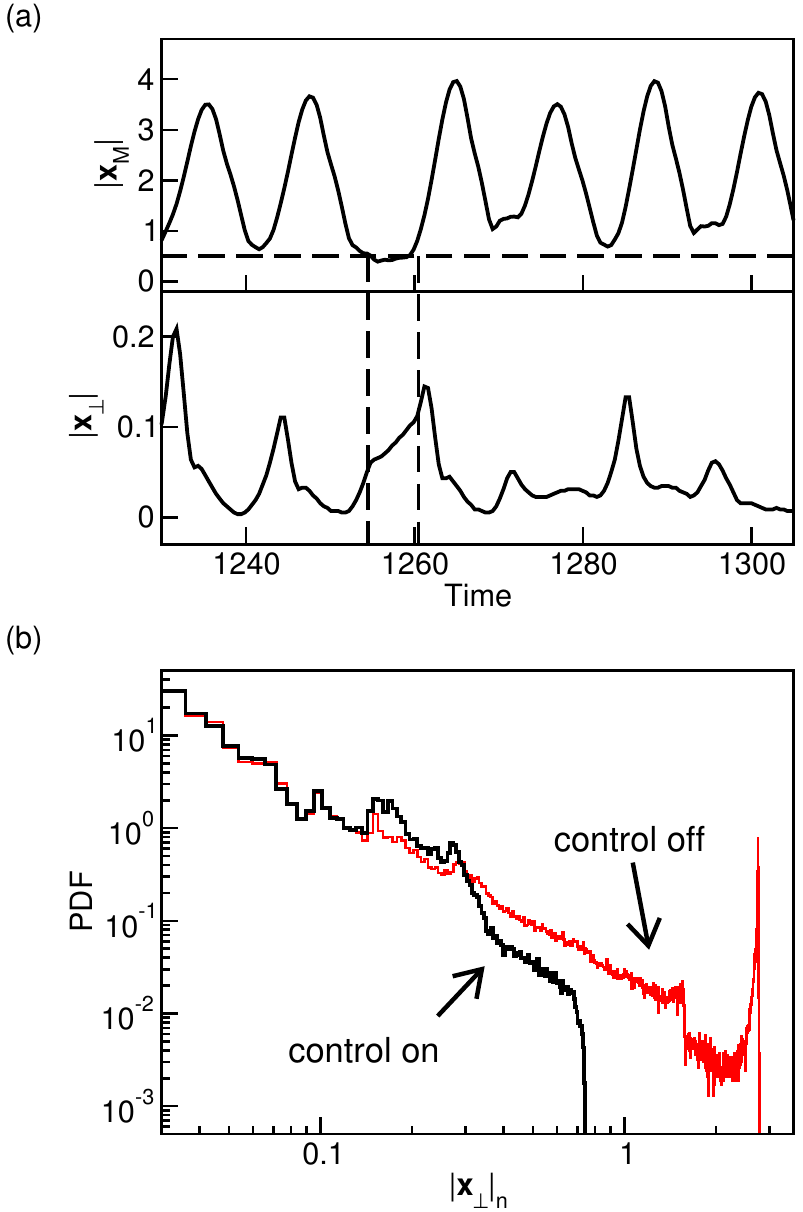}
\caption{(color online). Slaying dragon-kings. (a) Temporal evolution of the trajectory projected onto the invariant manifold ($\abs{\vect{x}_{M}}$) and the transverse space ($\abs{\vect{x}_{\perp}}$) during suppressed attractor bubbling. When $\abs{\vect{x}_{M}}$ is below $\abs{\vect{x}_{M}}_\mathrm{th} = 0.50$ (the horizontal dashed line) the occasional feedback is activated, reducing the height of a bubble (in the time interval between the two vertical dashed lines) that would grow large otherwise. (b) Probability density function for event-size $\abs{\vect{x}_{\perp}}_n$  in the presence (black) and absence (red online) of occasional proportional feedback. Here $c$ = 4.4, $c_{DK}$ = 0.55, $\vect{K}_{ij} = 1$ for $i=j=2$ and $0$ otherwise, and $\left(\vect{K}_{DK}\right)_{ij} = 1$ for $i=j=1$ and $0$ otherwise. A more detailed comparison between experiment and theory is presented in \cite{Note1}. 
\label{fig:slaying_DK}}
\end{figure}
Figure\ \ref{fig:slaying_DK}(a) shows the temporal evolution of the system in the presence of occasional feedback. When $\abs{\vect{x}_{M}}>\abs{\vect{x}_{M}}_\mathrm{th}$, no feedback is applied and the small bubbling events are allowed to proceed. On the other hand, when  $\abs{\vect{x}_{M}}<\abs{\vect{x}_{M}}_\mathrm{th}$, feedback perturbations are applied that are only 3\% of the system size (defined as the maximum value of $\abs{\vect{x}_{M}} \sim 4$).  Such small perturbation only causes a small change in $\vect{x}_S$, yet it has a dramatic change in $\vect{x}_\perp$:  the large bubble is suppressed. Over a long time scale, feedback is only applied 1.5\% of the time, consistent with the frequency and duration of extreme events. Thus, the total perturbation size averaged over the whole time, including the intervals when the perturbation is not active, corresponds to 0.05\% of the system size. As a result of this occasional feedback, we observe that the largest events, i
 ncluding the DKs, are entirely suppressed, as shown in the probability density function for $\abs{\vect{x}_{\perp}}$  in Fig.\ \ref{fig:slaying_DK}(b). It is seen that the small- to intermediate-size bubbles are unaffected; only the events that would have a large size in the absence of control are suppressed.

Our work addresses several important questions regarding complex systems.  We answer affirmatively and conclusively that: 1) a particular simple, but nontrivial system displays DKs whose event-size distribution deviates significantly upward from a power law in the tail; 2) DKs can be predicted; and 3) this predictability can be used to occasionally and efficiently activate countermeasures that suppress or mitigate the effects of DKs.  An important and immediate open question is whether it is possible to easily identify the unstable sets that are primarily responsible for causing DKs in the wide variety of complex systems that are already known to have attractor bubbling or in systems that may display bubbling but it is not yet appreciated that the behavior is of this type. 
While a specific method that is valid in all cases is unlikely to exist, the particular example studied here demonstrates that, with some understanding of the burst mechanism, large DK-type events may potentially be avoidable by devising small, well-chosen system perturbations.
Key to addressing this problem is the development of new tools for analyzing models of complex systems or for time series analysis of natural systems that can identify 
 burst mechanisms. 
We suggest that the use of this knowledge to devise appropriate control strategies is a worthy pursuit given the increasing appearance of extreme events and their impact on society.  

\begin{acknowledgments}
HLDSC and MO acknowledge the financial support from the Brazilian agencies Conselho Nacional de Desenvolvimento Cient\'{\i}fico e Tecnol\'ogico (CNPq) and Financiadora de Estudos e Projetos (FINEP). DJG gratefully acknowledges the financial support of the U.S. Office of Naval Research, 
grant \# N00014-07-1-0734 and thanks Joshua Bienfang for constructing the chaotic electronic circuits. EO and DJG gratefully acknowledge the financial support of the US ARO through grant W911NF-12-1-0101 and W911NF-12-1-0099, respectively.  The authors thank one of the anonymous referees for suggestions incorporated in this Letter.
\end{acknowledgments}

\bibliography{dragon_kings_refs}

\begin{thebibliography}{30}%
\makeatletter
\providecommand \@ifxundefined [1]{%
 \@ifx{#1\undefined}
}%
\providecommand \@ifnum [1]{%
 \ifnum #1\expandafter \@firstoftwo
 \else \expandafter \@secondoftwo
 \fi
}%
\providecommand \@ifx [1]{%
 \ifx #1\expandafter \@firstoftwo
 \else \expandafter \@secondoftwo
 \fi
}%
\providecommand \natexlab [1]{#1}%
\providecommand \enquote  [1]{``#1''}%
\providecommand \bibnamefont  [1]{#1}%
\providecommand \bibfnamefont [1]{#1}%
\providecommand \citenamefont [1]{#1}%
\providecommand \href@noop [0]{\@secondoftwo}%
\providecommand \href [0]{\begingroup \@sanitize@url \@href}%
\providecommand \@href[1]{\@@startlink{#1}\@@href}%
\providecommand \@@href[1]{\endgroup#1\@@endlink}%
\providecommand \@sanitize@url [0]{\catcode `\\12\catcode `\$12\catcode
  `\&12\catcode `\#12\catcode `\^12\catcode `\_12\catcode `\%12\relax}%
\providecommand \@@startlink[1]{}%
\providecommand \@@endlink[0]{}%
\providecommand \url  [0]{\begingroup\@sanitize@url \@url }%
\providecommand \@url [1]{\endgroup\@href {#1}{\urlprefix }}%
\providecommand \urlprefix  [0]{URL }%
\providecommand \Eprint [0]{\href }%
\providecommand \doibase [0]{http://dx.doi.org/}%
\providecommand \selectlanguage [0]{\@gobble}%
\providecommand \bibinfo  [0]{\@secondoftwo}%
\providecommand \bibfield  [0]{\@secondoftwo}%
\providecommand \translation [1]{[#1]}%
\providecommand \BibitemOpen [0]{}%
\providecommand \bibitemStop [0]{}%
\providecommand \bibitemNoStop [0]{.\EOS\space}%
\providecommand \EOS [0]{\spacefactor3000\relax}%
\providecommand \BibitemShut  [1]{\csname bibitem#1\endcsname}%
\let\auto@bib@innerbib\@empty
\bibitem [{\citenamefont {Nott}(2006)}]{Nott2006}%
  \BibitemOpen
  \bibfield  {author} {\bibinfo {author} {\bibfnamefont {J.}~\bibnamefont
  {Nott}},\ }\href@noop {} {\emph {\bibinfo {title} {Extreme Events: a Physical
  Reconstruction and Risk Assessment}}}\ (\bibinfo  {publisher} {Cambridge
  Univ. Press},\ \bibinfo {year} {2006})\BibitemShut {NoStop}%
\bibitem [{\citenamefont {Comfort}\ \emph {et~al.}(2010)\citenamefont
  {Comfort}, \citenamefont {Boin},\ and\ \citenamefont
  {Demchak}}]{Comfort2010}%
  \BibitemOpen
  \bibfield  {author} {\bibinfo {author} {\bibfnamefont {L.~K.}\ \bibnamefont
  {Comfort}}, \bibinfo {author} {\bibfnamefont {A.}~\bibnamefont {Boin}}, \
  and\ \bibinfo {author} {\bibfnamefont {C.~C.}\ \bibnamefont {Demchak}},\
  }\href@noop {} {\emph {\bibinfo {title} {Designing Resilience: Preparing for
  Extreme Events}}}\ (\bibinfo  {publisher} {Univ. of Pittsburgh Press},\
  \bibinfo {year} {2010})\BibitemShut {NoStop}%
\bibitem [{\citenamefont {Field}\ \emph {et~al.}(2012)\citenamefont {Field},
  \citenamefont {Barros}, \citenamefont {Stocker},\ and\ \citenamefont
  {Dahe}}]{Field2012}%
  \BibitemOpen
  \bibfield  {author} {\bibinfo {author} {\bibfnamefont {C.~B.}\ \bibnamefont
  {Field}}, \bibinfo {author} {\bibfnamefont {V.}~\bibnamefont {Barros}},
  \bibinfo {author} {\bibfnamefont {T.~F.}\ \bibnamefont {Stocker}}, \ and\
  \bibinfo {author} {\bibfnamefont {Q.}~\bibnamefont {Dahe}},\ }\href@noop {}
  {\emph {\bibinfo {title} {Managing the Risks of Extreme Events and Disasters
  to Advance Climate Change Adaptation}}}\ (\bibinfo  {publisher} {Cambridge
  University Press},\ \bibinfo {year} {2012})\BibitemShut {NoStop}%
\bibitem [{\citenamefont {Barnosky}\ \emph {et~al.}(2012)\citenamefont
  {Barnosky} \emph {et~al.}}]{Barnosky2012}%
  \BibitemOpen
  \bibfield  {author} {\bibinfo {author} {\bibfnamefont {A.~D.}\ \bibnamefont
  {Barnosky}} \emph {et~al.},\ }\href@noop {} {\bibfield  {journal} {\bibinfo
  {journal} {Nature}\ }\textbf {\bibinfo {volume} {486}},\ \bibinfo {pages}
  {52} (\bibinfo {year} {2012})}\BibitemShut {NoStop}%
\bibitem [{\citenamefont {Schwab}\ \emph {et~al.}(2013)\citenamefont {Schwab}
  \emph {et~al.}}]{GlobalRisks2013}%
  \BibitemOpen
  \bibfield  {author} {\bibinfo {author} {\bibfnamefont {K.}~\bibnamefont
  {Schwab}} \emph {et~al.},\ }\href@noop {} {\emph {\bibinfo {title} {Global
  Risks}}},\ \bibinfo {edition} {eighth}\ ed.\ (\bibinfo  {publisher} {World
  Economic Forum},\ \bibinfo {year} {2013})\BibitemShut {NoStop}%
\bibitem [{\citenamefont {Rundle}\ \emph {et~al.}(1996)\citenamefont {Rundle},
  \citenamefont {Turcotte},\ and\ \citenamefont {Klein}}]{Rundle1996}%
  \BibitemOpen
  \bibfield  {author} {\bibinfo {author} {\bibfnamefont {J.~B.}\ \bibnamefont
  {Rundle}}, \bibinfo {author} {\bibfnamefont {D.~L.}\ \bibnamefont
  {Turcotte}}, \ and\ \bibinfo {author} {\bibfnamefont {W.}~\bibnamefont
  {Klein}},\ }in\ \href@noop {} {\emph {\bibinfo {booktitle} {Reduction and
  predictability of natural disasters}}},\ \bibinfo {series} {Studies in the
  Science of Complexity}, Vol.\ \bibinfo {volume} {XXV}\ (\bibinfo  {publisher}
  {Addison-Wesley},\ \bibinfo {year} {1996})\BibitemShut {NoStop}%
\bibitem [{\citenamefont {Sornette}(2002)}]{Sornette2002}%
  \BibitemOpen
  \bibfield  {author} {\bibinfo {author} {\bibfnamefont {D.}~\bibnamefont
  {Sornette}},\ }\href@noop {} {\bibfield  {journal} {\bibinfo  {journal}
  {Proc. Natl. Acad. Sci. USA}\ }\textbf {\bibinfo {volume} {99}},\ \bibinfo
  {pages} {2522} (\bibinfo {year} {2002})}\BibitemShut {NoStop}%
\bibitem [{\citenamefont {Albeverio}\ \emph {et~al.}(2005)\citenamefont
  {Albeverio}, \citenamefont {Jentsch},\ and\ \citenamefont
  {Kantz}}]{Albeverio2005}%
  \BibitemOpen
  \bibfield  {author} {\bibinfo {author} {\bibfnamefont {S.}~\bibnamefont
  {Albeverio}}, \bibinfo {author} {\bibfnamefont {V.}~\bibnamefont {Jentsch}},
  \ and\ \bibinfo {author} {\bibfnamefont {H.}~\bibnamefont {Kantz}},\
  }\href@noop {} {\emph {\bibinfo {title} {Extreme Events in Nature and
  Society}}},\ \bibinfo {series} {The Frontiers Collection}, Vol.\ \bibinfo
  {volume} {XVI}\ (\bibinfo  {publisher} {Springer},\ \bibinfo {year}
  {2005})\BibitemShut {NoStop}%
\bibitem [{\citenamefont {Scheffer}\ \emph {et~al.}(2012)\citenamefont
  {Scheffer} \emph {et~al.}}]{Scheffer2012}%
  \BibitemOpen
  \bibfield  {author} {\bibinfo {author} {\bibfnamefont {M.}~\bibnamefont
  {Scheffer}} \emph {et~al.},\ }\href@noop {} {\bibfield  {journal} {\bibinfo
  {journal} {Science}\ }\textbf {\bibinfo {volume} {338}},\ \bibinfo {pages}
  {344} (\bibinfo {year} {2012})}\BibitemShut {NoStop}%
\bibitem [{\citenamefont {Biggs}\ \emph {et~al.}(2009)\citenamefont {Biggs},
  \citenamefont {Carpenter},\ and\ \citenamefont {Brock}}]{Biggs2009}%
  \BibitemOpen
  \bibfield  {author} {\bibinfo {author} {\bibfnamefont {R.}~\bibnamefont
  {Biggs}}, \bibinfo {author} {\bibfnamefont {S.~R.}\ \bibnamefont
  {Carpenter}}, \ and\ \bibinfo {author} {\bibfnamefont {W.~A.}\ \bibnamefont
  {Brock}},\ }\href@noop {} {\bibfield  {journal} {\bibinfo  {journal} {Proc.
  Natl. Acad. Sci. USA}\ }\textbf {\bibinfo {volume} {106}},\ \bibinfo {pages}
  {826} (\bibinfo {year} {2009})}\BibitemShut {NoStop}%
\bibitem [{\citenamefont {Dai}\ \emph {et~al.}(2012)\citenamefont {Dai},
  \citenamefont {Vorselen}, \citenamefont {Korolev},\ and\ \citenamefont
  {Gore}}]{Dai2012}%
  \BibitemOpen
  \bibfield  {author} {\bibinfo {author} {\bibfnamefont {L.}~\bibnamefont
  {Dai}}, \bibinfo {author} {\bibfnamefont {D.}~\bibnamefont {Vorselen}},
  \bibinfo {author} {\bibfnamefont {K.~S.}\ \bibnamefont {Korolev}}, \ and\
  \bibinfo {author} {\bibfnamefont {J.}~\bibnamefont {Gore}},\ }\href@noop {}
  {\bibfield  {journal} {\bibinfo  {journal} {Science}\ }\textbf {\bibinfo
  {volume} {336}},\ \bibinfo {pages} {1175} (\bibinfo {year}
  {2012})}\BibitemShut {NoStop}%
\bibitem [{\citenamefont {Krawiecki}\ \emph {et~al.}(2002)\citenamefont
  {Krawiecki}, \citenamefont {Holyst},\ and\ \citenamefont
  {Helbing}}]{Krawiecki2002}%
  \BibitemOpen
  \bibfield  {author} {\bibinfo {author} {\bibfnamefont {A.}~\bibnamefont
  {Krawiecki}}, \bibinfo {author} {\bibfnamefont {J.~A.}\ \bibnamefont
  {Holyst}}, \ and\ \bibinfo {author} {\bibfnamefont {D.}~\bibnamefont
  {Helbing}},\ }\href@noop {} {\bibfield  {journal} {\bibinfo  {journal} {Phys.
  Rev. Lett.}\ }\textbf {\bibinfo {volume} {89}},\ \bibinfo {pages} {158701}
  (\bibinfo {year} {2002})}\BibitemShut {NoStop}%
\bibitem [{\citenamefont {Sornette}(2009)}]{Sornette2009}%
  \BibitemOpen
  \bibfield  {author} {\bibinfo {author} {\bibfnamefont {D.}~\bibnamefont
  {Sornette}},\ }\href@noop {} {\bibfield  {journal} {\bibinfo  {journal}
  {Intl. J. Terraspace Sci. Eng.}\ }\textbf {\bibinfo {volume} {2}},\ \bibinfo
  {pages} {1} (\bibinfo {year} {2009})}\BibitemShut {NoStop}%
\bibitem [{\citenamefont {Taleb}(2007)}]{Taleb2007}%
  \BibitemOpen
  \bibfield  {author} {\bibinfo {author} {\bibfnamefont {N.~N.}\ \bibnamefont
  {Taleb}},\ }\href@noop {} {\emph {\bibinfo {title} {The Black Swan: The
  Impact of the Highly Improbable}}}\ (\bibinfo  {publisher} {Random House},\
  \bibinfo {year} {2007})\BibitemShut {NoStop}%
\bibitem [{\citenamefont {Bak}(1996)}]{Bak1996}%
  \BibitemOpen
  \bibfield  {author} {\bibinfo {author} {\bibfnamefont {P.}~\bibnamefont
  {Bak}},\ }\href@noop {} {\emph {\bibinfo {title} {How Nature Works: The
  Science of Self-Organized Criticality}}}\ (\bibinfo  {publisher} {Springer},\
  \bibinfo {year} {1996})\BibitemShut {NoStop}%
\bibitem [{\citenamefont {Embrechts}\ \emph {et~al.}(2011)\citenamefont
  {Embrechts}, \citenamefont {Kl�ppelberg},\ and\ \citenamefont
  {Mikosch}}]{Embrechts2011}%
  \BibitemOpen
  \bibfield  {author} {\bibinfo {author} {\bibfnamefont {P.}~\bibnamefont
  {Embrechts}}, \bibinfo {author} {\bibfnamefont {C.}~\bibnamefont
  {Kl�ppelberg}}, \ and\ \bibinfo {author} {\bibfnamefont {T.}~\bibnamefont
  {Mikosch}},\ }\href@noop {} {\emph {\bibinfo {title} {Modelling Extremal
  Events for Insurance and Finance}}},\ \bibinfo {edition} {corrected}\ ed.\
  (\bibinfo  {publisher} {Springer},\ \bibinfo {address} {Heidelberg},\
  \bibinfo {year} {2011})\BibitemShut {NoStop}%
\bibitem [{\citenamefont {Knight}(1921)}]{Knight1921}%
  \BibitemOpen
  \bibfield  {author} {\bibinfo {author} {\bibfnamefont {F.~H.}\ \bibnamefont
  {Knight}},\ }\href@noop {} {\emph {\bibinfo {title} {Risk, Uncertainty, and
  Profit}}}\ (\bibinfo  {publisher} {Houghton Mifflin Co.},\ \bibinfo {address}
  {Boston, MA},\ \bibinfo {year} {1921})\BibitemShut {NoStop}%
\bibitem [{\citenamefont {Sornette}\ and\ \citenamefont
  {Ouillon}(2012)}]{Sornette2012}%
  \BibitemOpen
  \bibfield  {author} {\bibinfo {author} {\bibfnamefont {D.}~\bibnamefont
  {Sornette}}\ and\ \bibinfo {author} {\bibfnamefont {G.}~\bibnamefont
  {Ouillon}},\ }\href@noop {} {\bibfield  {journal} {\bibinfo  {journal} {Eur.
  Phys. J. Special Topics}\ }\textbf {\bibinfo {volume} {25}},\ \bibinfo
  {pages} {1} (\bibinfo {year} {2012})}\BibitemShut {NoStop}%
\bibitem [{\citenamefont {Schmittbuhl}\ \emph {et~al.}(1993)\citenamefont
  {Schmittbuhl}, \citenamefont {Vilotte},\ and\ \citenamefont
  {Roux}}]{Schmittbuhl1993}%
  \BibitemOpen
  \bibfield  {author} {\bibinfo {author} {\bibfnamefont {J.}~\bibnamefont
  {Schmittbuhl}}, \bibinfo {author} {\bibfnamefont {J.-P.}\ \bibnamefont
  {Vilotte}}, \ and\ \bibinfo {author} {\bibfnamefont {S.}~\bibnamefont
  {Roux}},\ }\href@noop {} {\bibfield  {journal} {\bibinfo  {journal}
  {Europhys. Lett.}\ }\textbf {\bibinfo {volume} {21}},\ \bibinfo {pages} {375}
  (\bibinfo {year} {1993})}\BibitemShut {NoStop}%
\bibitem [{\citenamefont {Gong}\ \emph {et~al.}(2007)\citenamefont {Gong},
  \citenamefont {Nikolaev},\ and\ \citenamefont {van Leeuwen}}]{Gong2007}%
  \BibitemOpen
  \bibfield  {author} {\bibinfo {author} {\bibfnamefont {P.}~\bibnamefont
  {Gong}}, \bibinfo {author} {\bibfnamefont {A.~R.}\ \bibnamefont {Nikolaev}},
  \ and\ \bibinfo {author} {\bibfnamefont {C.}~\bibnamefont {van Leeuwen}},\
  }\href@noop {} {\bibfield  {journal} {\bibinfo  {journal} {Phys. Rev. E}\
  }\textbf {\bibinfo {volume} {76}},\ \bibinfo {pages} {011904} (\bibinfo
  {year} {2007})}\BibitemShut {NoStop}%
\bibitem [{\citenamefont {Takayasu}\ \emph {et~al.}(1992)\citenamefont
  {Takayasu}, \citenamefont {Miura}, \citenamefont {Hirabayashi},\ and\
  \citenamefont {Hamada}}]{Takayasu1992}%
  \BibitemOpen
  \bibfield  {author} {\bibinfo {author} {\bibfnamefont {H.}~\bibnamefont
  {Takayasu}}, \bibinfo {author} {\bibfnamefont {H.}~\bibnamefont {Miura}},
  \bibinfo {author} {\bibfnamefont {T.}~\bibnamefont {Hirabayashi}}, \ and\
  \bibinfo {author} {\bibfnamefont {K.}~\bibnamefont {Hamada}},\ }\href@noop {}
  {\bibfield  {journal} {\bibinfo  {journal} {Physica A}\ }\textbf {\bibinfo
  {volume} {184}},\ \bibinfo {pages} {127} (\bibinfo {year}
  {1992})}\BibitemShut {NoStop}%
\bibitem [{\citenamefont {Ott}(2002)}]{Ott2002}%
  \BibitemOpen
  \bibfield  {author} {\bibinfo {author} {\bibfnamefont {E.}~\bibnamefont
  {Ott}},\ }\href@noop {} {\emph {\bibinfo {title} {Chaos in Dynamical
  Systems}}},\ \bibinfo {edition} {2nd}\ ed.\ (\bibinfo  {publisher} {Cambridge
  Univ. Press},\ \bibinfo {address} {New York},\ \bibinfo {year}
  {2002})\BibitemShut {NoStop}%
\bibitem [{\citenamefont {Sommerer}\ and\ \citenamefont
  {Ott}(1993)}]{Sommerer1993}%
  \BibitemOpen
  \bibfield  {author} {\bibinfo {author} {\bibfnamefont {J.~C.}\ \bibnamefont
  {Sommerer}}\ and\ \bibinfo {author} {\bibfnamefont {E.}~\bibnamefont {Ott}},\
  }\href@noop {} {\bibfield  {journal} {\bibinfo  {journal} {Nature}\ }\textbf
  {\bibinfo {volume} {365}},\ \bibinfo {pages} {138} (\bibinfo {year}
  {1993})}\BibitemShut {NoStop}%
\bibitem [{\citenamefont {Mosekilde}\ \emph {et~al.}(2002)\citenamefont
  {Mosekilde}, \citenamefont {Postnov},\ and\ \citenamefont
  {Maistrenko}}]{Mosekilde2002}%
  \BibitemOpen
  \bibfield  {author} {\bibinfo {author} {\bibfnamefont {E.}~\bibnamefont
  {Mosekilde}}, \bibinfo {author} {\bibfnamefont {D.}~\bibnamefont {Postnov}},
  \ and\ \bibinfo {author} {\bibfnamefont {Y.}~\bibnamefont {Maistrenko}},\
  }\href@noop {} {\emph {\bibinfo {title} {Chaotic Synchronization:
  Applications to Living Systems}}},\ \bibinfo {series} {Nonlinear Science},
  Vol.~\bibinfo {volume} {42}\ (\bibinfo  {publisher} {World Scientific},\
  \bibinfo {year} {2002})\BibitemShut {NoStop}%
\bibitem [{\citenamefont {Ashwin}\ \emph {et~al.}(1994)\citenamefont {Ashwin},
  \citenamefont {Buescu},\ and\ \citenamefont {Stewart}}]{Ashwin1994}%
  \BibitemOpen
  \bibfield  {author} {\bibinfo {author} {\bibfnamefont {P.}~\bibnamefont
  {Ashwin}}, \bibinfo {author} {\bibfnamefont {J.}~\bibnamefont {Buescu}}, \
  and\ \bibinfo {author} {\bibfnamefont {I.}~\bibnamefont {Stewart}},\
  }\href@noop {} {\bibfield  {journal} {\bibinfo  {journal} {Phys. Lett. A}\
  }\textbf {\bibinfo {volume} {193}},\ \bibinfo {pages} {126} (\bibinfo {year}
  {1994})}\BibitemShut {NoStop}%
\bibitem [{Note1()}]{Note1}%
  \BibitemOpen
  \bibinfo {note} {See Supplemental Material at [URL will be inserted by
  publisher] for additional figures and details.}\BibitemShut {Stop}%
\bibitem [{\citenamefont {Gauthier}\ and\ \citenamefont
  {Bienfang}(1996)}]{Gauthier1996}%
  \BibitemOpen
  \bibfield  {author} {\bibinfo {author} {\bibfnamefont {D.~J.}\ \bibnamefont
  {Gauthier}}\ and\ \bibinfo {author} {\bibfnamefont {J.~C.}\ \bibnamefont
  {Bienfang}},\ }\href@noop {} {\bibfield  {journal} {\bibinfo  {journal}
  {Phys. Rev. Lett.}\ }\textbf {\bibinfo {volume} {77}},\ \bibinfo {pages}
  {1751} (\bibinfo {year} {1996})}\BibitemShut {NoStop}%
\bibitem [{\citenamefont {Heagy}\ \emph {et~al.}(1995)\citenamefont {Heagy},
  \citenamefont {Carroll},\ and\ \citenamefont {Pecora}}]{Heagy1995}%
  \BibitemOpen
  \bibfield  {author} {\bibinfo {author} {\bibfnamefont {J.~F.}\ \bibnamefont
  {Heagy}}, \bibinfo {author} {\bibfnamefont {T.~L.}\ \bibnamefont {Carroll}},
  \ and\ \bibinfo {author} {\bibfnamefont {L.~M.}\ \bibnamefont {Pecora}},\
  }\href@noop {} {\bibfield  {journal} {\bibinfo  {journal} {Phys. Rev. E}\
  }\textbf {\bibinfo {volume} {52}},\ \bibinfo {pages} {R1253} (\bibinfo {year}
  {1995})}\BibitemShut {NoStop}%
\bibitem [{\citenamefont {Venkataramani}\ \emph {et~al.}(1996)\citenamefont
  {Venkataramani}, \citenamefont {Hunt}, \citenamefont {Ott}, \citenamefont
  {Gauthier},\ and\ \citenamefont {Bienfang}}]{Venkataramani1996}%
  \BibitemOpen
  \bibfield  {author} {\bibinfo {author} {\bibfnamefont {S.~C.}\ \bibnamefont
  {Venkataramani}}, \bibinfo {author} {\bibfnamefont {B.~R.}\ \bibnamefont
  {Hunt}}, \bibinfo {author} {\bibfnamefont {E.}~\bibnamefont {Ott}}, \bibinfo
  {author} {\bibfnamefont {D.~J.}\ \bibnamefont {Gauthier}}, \ and\ \bibinfo
  {author} {\bibfnamefont {J.~C.}\ \bibnamefont {Bienfang}},\ }\href@noop {}
  {\bibfield  {journal} {\bibinfo  {journal} {Phys. Rev. Lett.}\ }\textbf
  {\bibinfo {volume} {77}},\ \bibinfo {pages} {5361} (\bibinfo {year}
  {1996})}\BibitemShut {NoStop}%
\bibitem [{\citenamefont {Newell}\ \emph {et~al.}(1994)\citenamefont {Newell},
  \citenamefont {Alsing}, \citenamefont {Gavrielides},\ and\ \citenamefont
  {Kovanis}}]{Newell1994}%
  \BibitemOpen
  \bibfield  {author} {\bibinfo {author} {\bibfnamefont {T.~C.}\ \bibnamefont
  {Newell}}, \bibinfo {author} {\bibfnamefont {P.~M.}\ \bibnamefont {Alsing}},
  \bibinfo {author} {\bibfnamefont {A.}~\bibnamefont {Gavrielides}}, \ and\
  \bibinfo {author} {\bibfnamefont {V.}~\bibnamefont {Kovanis}},\ }\href@noop
  {} {\bibfield  {journal} {\bibinfo  {journal} {Phys. Rev. Lett.}\ }\textbf
  {\bibinfo {volume} {72}},\ \bibinfo {pages} {1647} (\bibinfo {year}
  {1994})}\BibitemShut {NoStop}%
\end{thebibliography}%

\end{document}



\title{Supplementary Material to \\ ``Predictability and suppression of extreme events in a chaotic system''}


\affiliation{Grupo de F\'{\i}sica At\^omica e Lasers - DF, Universidade Federal da Para\'{\i}ba, Caixa Postal 5086 - 58051-900 - Jo\~{a}o Pessoa, PB - Brazil.}

\author{Hugo L. D. de S. Cavalcante}
\author{Marcos Ori\'{a}}
\affiliation{Grupo de F\'{\i}sica At\^omica e Lasers - DF, Universidade Federal da Para\'{\i}ba, Caixa Postal 5086 - 58051-900 - Jo\~{a}o Pessoa, PB - Brazil.}

\author{Didier Sornette}
\affiliation{ETH Zurich, Department of Management, Technology and Economics, Scheuchzerstrasse 7, CH-8092 Zurich, Switzerland.}

\author{Edward Ott}
\affiliation{Institute for Research in Electronics and Applied Physics, Institute for Systems Research, University of Maryland, College Park, MD, 20742 USA.}

\author{Daniel J. Gauthier}
\affiliation{Department of Physics, Duke University, Box 90305, Durham, NC, 27708 USA.}





\maketitle

\section{Chaotic Electronic Circuit}

\begin{figure}[!tbp]
\resizebox{8.5cm}{!}{\includegraphics{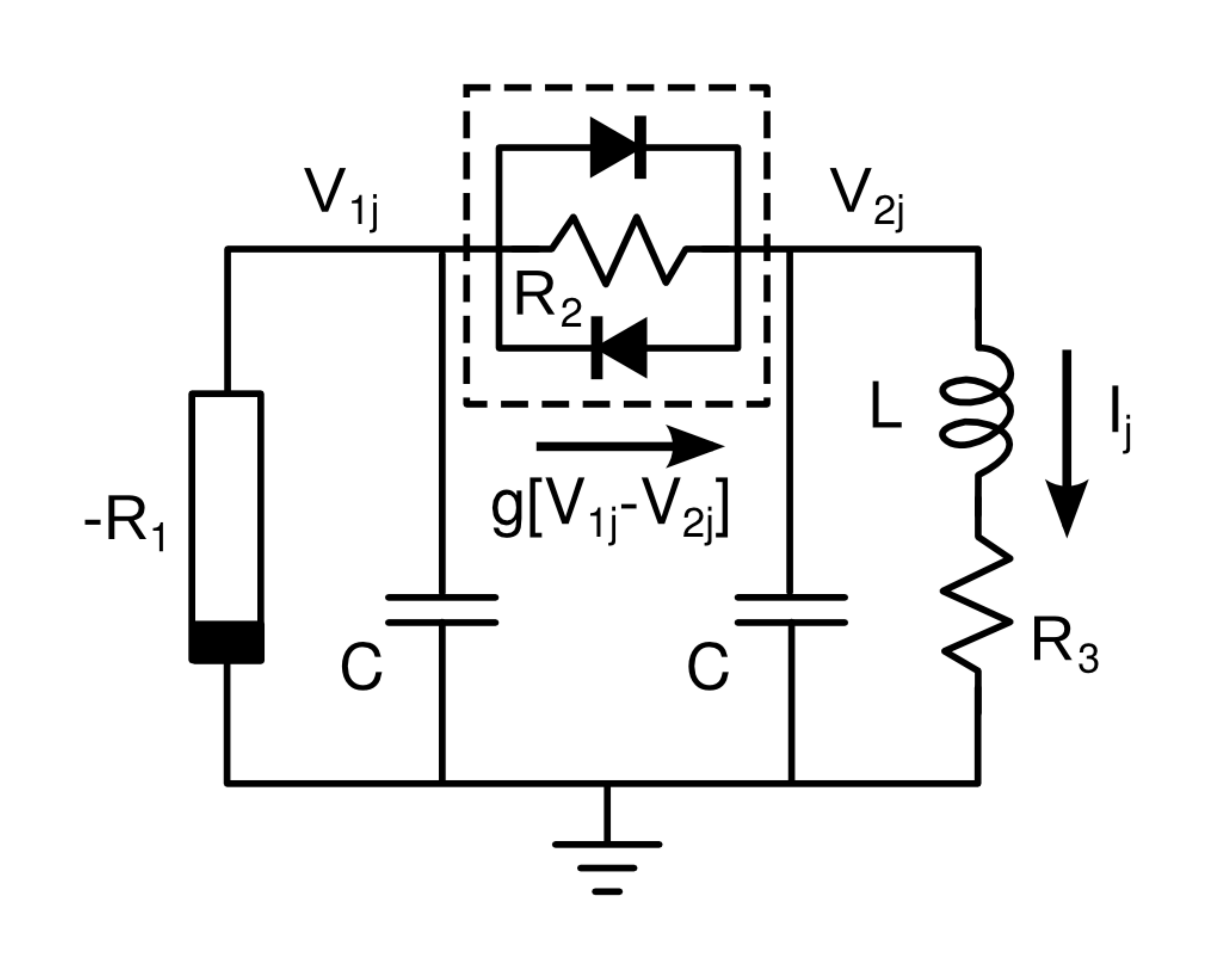}}
\caption{{\bf Chaotic electronic oscillator} with negative resistance -2,814 $\Omega$, 10 nF capacitors, 55 mH inductor (dc resistance 353 $\Omega$) in series with resistance 100 $\Omega$, and a passive nonlinear element (resistance 8,067 $\Omega$, and back-to-back parallel diodes, type 1N914, dashed box). \label{fig:circuit}}
\end{figure}

A single chaotic electronic oscillator used in our study is identical to that used by Gauthier and Bienfang in their study of attractor bubbling [27]. Here, we summarize the properties of the circuit for completeness, which consists of a negative resistor wired in series with a capacitor, which is coupled to an inductor-resistor-capacitor tank circuit through a nonlinear conductance, as shown schematically in Fig.\ \ref{fig:circuit}. Its behavior is governed by the set of dimensionless equations (3-5) in the Letter, where $V_{1j}$ and $V_{2j}$ represent the voltage drop across the capacitors (normalized to the diode voltage $V_{d}=$ 0.58 V), $I_{j}$ represents the current flowing through the inductor (normalized to $I_{d} = V_d /R = $ 0.25 mA for $R =$ 2,345 $\Omega$), $g\left[V\right] = V/R_2 +I_r \left( \exp(\alpha_f V)-\exp(-\alpha_r V) \right)$ (Eq. (6)) represents the current (normalized to $I_d$) flowing through the parallel combination of the resistor and diodes, and t
 ime is normalized to $\tau = \sqrt{LC} = $ 23.5 $\mu$s. The circuit displays ``double scroll'' behavior for $I_r$= 22.5 $\mu$A, $\alpha_f = \alpha_r= $ 11.6, $R_1 =$ 1.30, $R_2 =$ 3.44, $R_3 =$ 0.043, $R_\text{dc} =$ 0.15 (the dc resistance of the inductor), and $R_4 = R_3+R_\text{dc} = $ 0.193, where all the resistances are normalized to $R$. The state vector is given in terms of these variables via the relation $\vect{x}^T_j = (V_{1j}, V_{2j}, I_j)$.
The chaotic electronic circuits are coupled by measuring the differences of either $V_{1j}$ or $V_{2j}$ in each circuit with an instrumentation amplifier, converting to a current, and injecting the resulting signal to the slave node above one of the capacitors.

\section{Local Lyapunov Exponents}
Attractor bubbling occurs when the largest transverse Lyapunov exponent $\lambda_{\bot} $ is negative (the synchronized state is stable in the average sense), but there exist positive local transverse Lyapunov exponents $\eta_{\bot}$ of invariant sets embedded in the chaotic attractor.  The transverse Lyapunov exponents $\lambda_{\bot} $ and $\eta_{\bot} $ are determined from the variational equation
\begin{equation}
\frac{d\delta\vect{x}_\bot}{dt} = \left\{D\vect{F}\left[\vect{s}(t) \right]-c\vect{K} \right\} \delta\vect{x}_\bot, \label{eq:transverse_flow} 
\end{equation}
where $D\vect{F}\left[\vect{s}(t) \right]$ is the Jacobian of $\vect{F}$ evaluated on $\vect{s}(t)$, and, for $\lambda_{\bot} $, $\vect{s}(t)$ is a typical orbit on the chaotic attractor, while, for $\eta_{\bot}$, $\vect{s}(t)$ is an ergodic orbit on the invariant set of interest embedded in the chaotic atractor. Determining the value of the largest Lyapunov exponent $\lambda_{\bot} $ (or $\eta_{\bot}$) involves integrating Eq.\ \eqref{eq:transverse_flow} for a short time, renormalizing $\delta\vect{x}_\bot$ to avoid computer overflow, and continuing the integration until the attractor (or the relevant invariant set) is largely covered. 
\begin{figure}[!tbhp]
\resizebox{8.5cm}{!}{\includegraphics{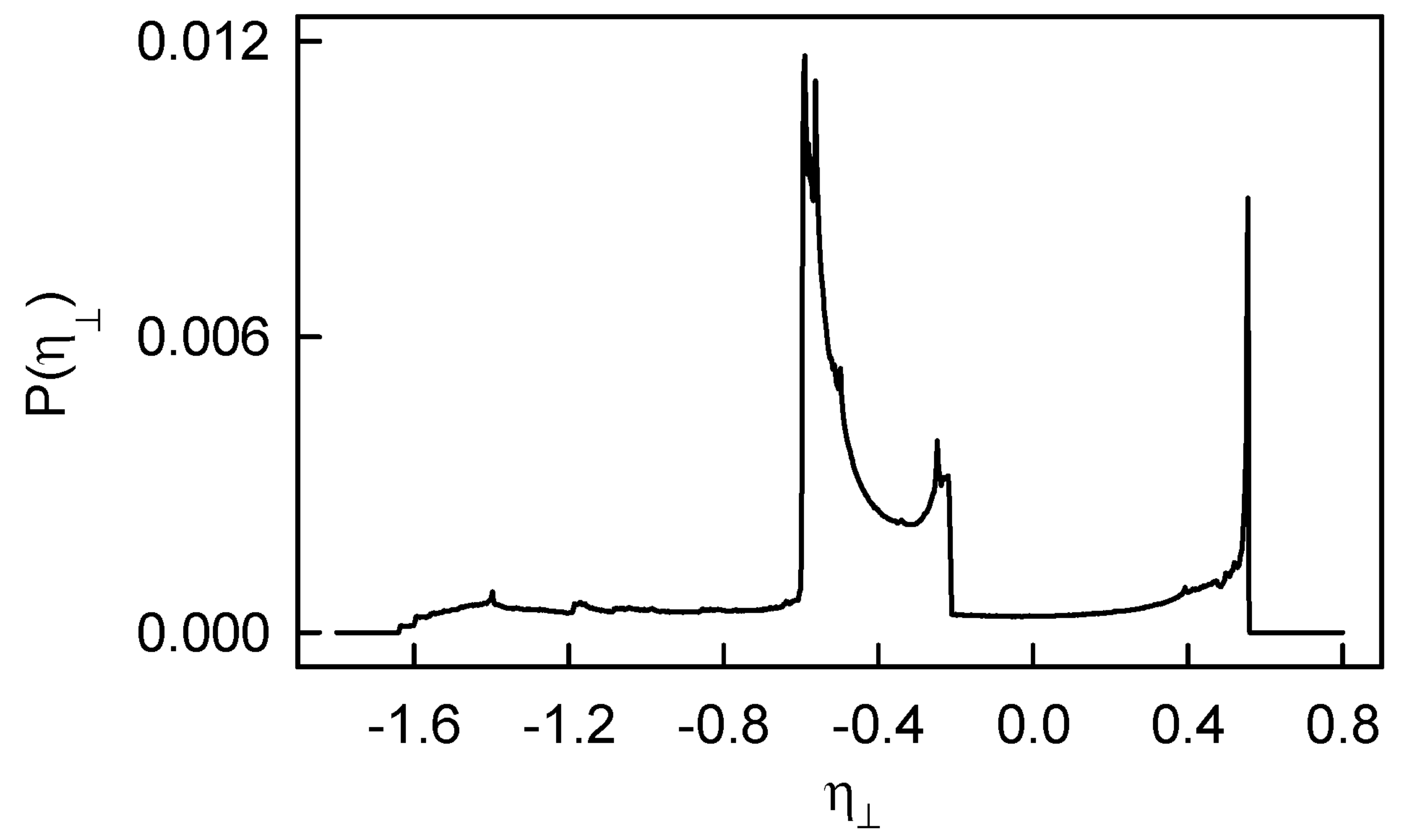}}
\caption{{\bf Probability distribution function of the short-time, or local Lyapunov exponents} for the coupled chaotic oscillators. \label{fig:local_lyap}}
\end{figure}

For the purpose of understanding bubbling, it is the largest local (short-time) Lyapunov exponents that are of interest.  It is the times when $\eta_\bot$ is positive that the trajectory $\vect{s}(t)$ is near an unstable set embedded in the attractor and a bubbling event is likely. Figure S2 shows the probability density for $\eta_\bot$ using a short integration time of 0.005 dimensionless time units. We see that the distribution is very non-Gaussian at short times, and, most importantly, is sharply peaked around $\eta_\bot \sim 0.55$.  Analysis shows that this large peak is associated with the unstable saddle embedded in the attractor at $\vect{x}_\bot = 0$.  It is this unstable saddle that causes trajectories to transition between the double scrolls of the attractor. 

The analysis so far assumes that the components of the master and slave electronic oscillators are identical and that there is no noise in the system.  We took great care to match all components to better than 1\%, and the noise level is small.  In this case, it is useful to suppose the existence of an effective approximate invariant manifold, but these nonideal behaviors (i.e., noise and mismatch) are required to observe bubbling. 
When, at some time $t_0$, the trajectory gets close to an ideal unstable set (such as the unstable saddle), we can regard the nonideal effects as acting like a transverse  initial perturbation, which subsequently leads the orbit to experience exponential growth according to
\begin{equation}
\left|\vect{x}_\bot\right| = \left|\vect{x}_\bot (t_0) \right| \exp\left[\eta_\bot (t-t_0) \right], \label{eq:transverse_solution} 
\end{equation}
for $t-t_0$ smaller than the characteristic time scale over which the trajectory remains within a neighborhood of the unstable set whose (local transverse) stability is given by $\eta_\bot$. This is the initiation of the bubble. 


\section{Theoretical Derivation of the Distribution }
\label{sec:theory}
\begin{figure}[!htbp]
\resizebox{8.5cm}{!}{\includegraphics{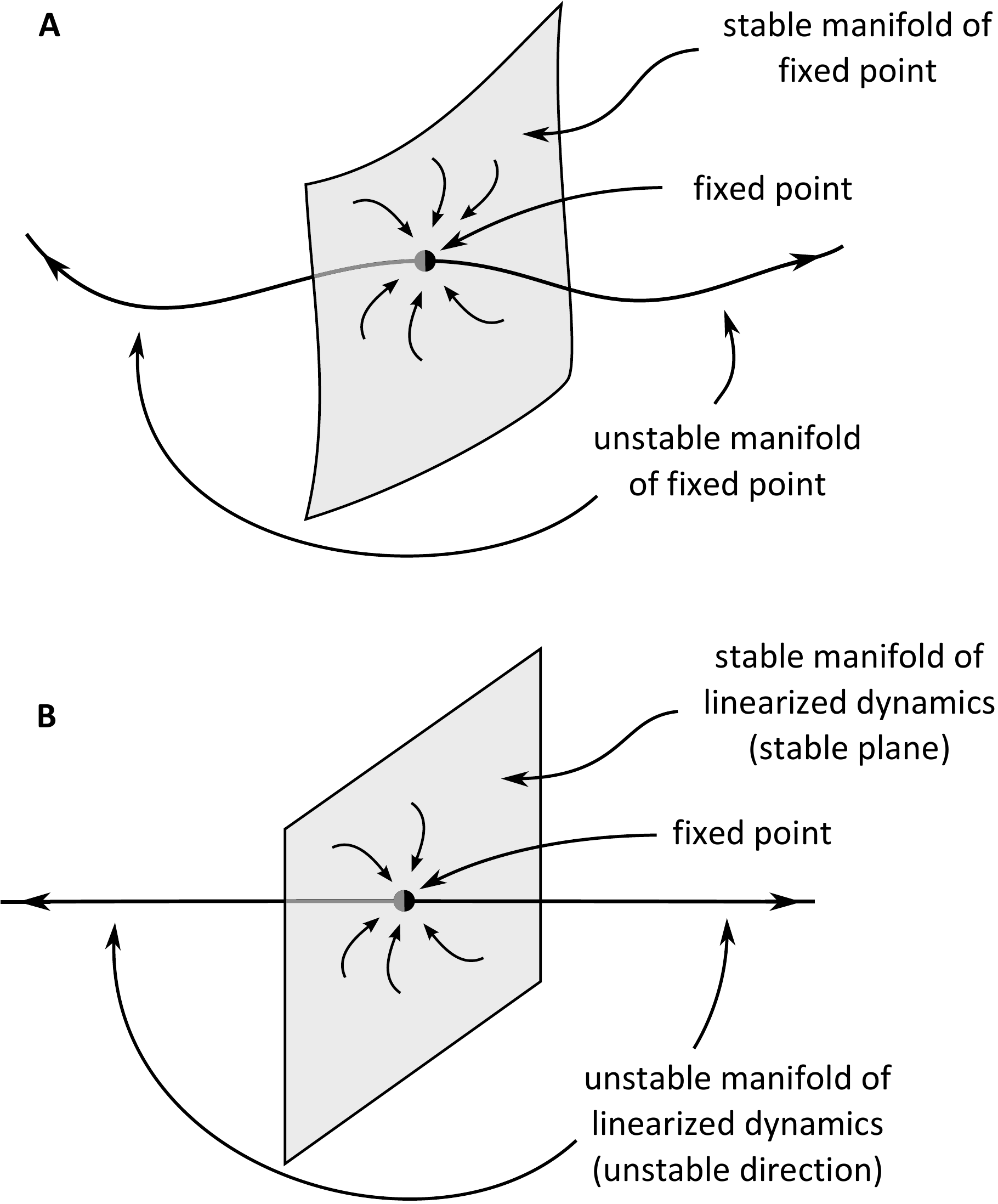}}
\caption{{\bf Manifolds of the fixed point} ({\bf A}) for the nonlinear flow and ({\bf B}) for the linearized dynamics. \label{fig:manifolds}}
\end{figure}

We now present a theoretical analysis for the observed power law exponent $-2$, and we also give an argument to explain why the distribution is truncated at a maximum event size and ends in a peak. 
In order to build the PDF of event sizes, let us follow the trajectories $\vect{x}_M(t)$ of the master and $\vect{x}_S(t)$ of the slave subsystems when they both  start, at time $t_0$, in a volume of phase space close to the unstable, saddle-type fixed point at the origin. 
If the distance to the saddle-point is small enough, the dynamics inside this volume can be approximated by linearizing the evolution equations (Eqs. (3-5)) around the saddle-point. The structure of the linearized phase space depends on the values of the characteristic exponents of the saddle, which are obtained as the eigenvalues of the 3-D Jacobians for the master ($D\vect{F}[0]$) and slave ($D\vect{F}[0]-c\vect{K}$) subsystems:
For the master subsystem, $\gamma_{1M} = 0.517$ is the positive exponent; $\gamma_{2M}$ and $\gamma_{3M}$ are complex conjugates with negative real part, $\gamma_{2M,3M}= -0.261 \pm 0.970 i$. For the slave subsystem, $\gamma_{1S} = 0.491$, $\gamma_{2S} = -0.422$, $\gamma_{3S} = -4.47$ are all real. 
Each subsystem $j = M,S$ has two eigenvalues with real part negative, $\gamma_{2j}$ and $\gamma_{3j}$, which define a  \emph{stable plane}, a linearization of the stable manifold of the nonlinear dynamics close to the saddle-point. Transverse to the stable plane, there is an \emph{unstable direction} that corresponds to the direction of the eigenvector associated to the positive eigenvalue $\gamma_{1j}$. The projection of each subsystem trajectory onto its respective stable plane decreases exponentially with an exponent $\Re\!e\left\{{\gamma_{2M}}\right\}$ for the master and at least $\gamma_{2S}$ for the slave. The projection of each subsystem trajectory along its respective unstable direction grows exponentially with exponent $\gamma_{1j}$.
This description is represented pictorially in Fig.\ \ref{fig:manifolds}.

\begin{figure}[!tbhp]
\resizebox{8.5cm}{!}{\includegraphics{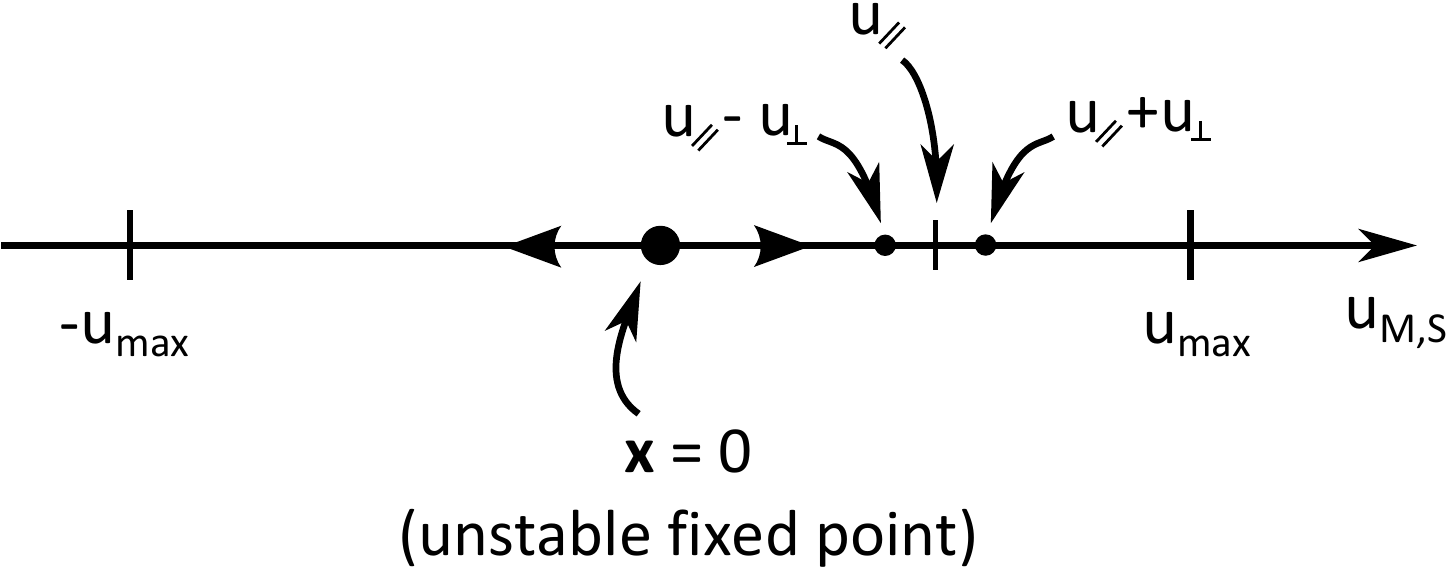}}
\caption{{\bf Linearized dynamics along the unstable direction.} \label{fig:unstable_direction}}
\end{figure}

Notice that the magnitudes of the unstable exponent $\gamma_{1j}$ of the master and slave subsystems are nearly equal (5\% difference). Thus, for simplicity, we take  both positive exponents equal, and define $\eta = \gamma_{1M} \cong \gamma_{1S}$. The 1-D projections of the trajectories $\vect{x}_M$ and  $\vect{x}_S$ along the unstable direction are denoted by $u_M$ and $u_S$, respectively. The dynamics along these 1-D axes are depicted in Fig.\ \ref{fig:unstable_direction}. Written in terms of their initial center $u_\parallelslant(t_0) = \left[u_M(t_0)+u_S(t_0)\right]/2$ and oriented half-separation $u_\bot(t_0) = \left[ u_M(t_0)-u_S(t_0)\right]/2$, these projections obey the solutions of the unstable part of the linearized equations 
\begin{align}
u_M(t) & = \left[u_\parallelslant(t_0)+u_\bot(t_0) \right] \exp\left[\eta(t-t_0)\right], \label{eq:u_M}\\ 
u_S(t) & = \left[u_\parallelslant(t_0)-u_\bot(t_0) \right] \exp\left[\eta(t-t_0)\right], \label{eq:u_S} 
\end{align}
and grow exponentially until the contribution of higher-order terms become significant in the dynamics and the linear approximation of the flow near the saddle-point is no longer valid. We call this latter phenomenon \emph{saturation} of the exponential growth. After saturation, the trajectories can still grow at a different, nonexponential rate, and eventually are brought back by the nonlinear dynamics to a region near the saddle, in a process ending the burst away from the attractor that we call \emph{reinjection}. We want to calculate the distribution of the burst sizes $B$, where $B$ is the maximum value of the separation between the trajectory and the invariant manifold during the burst.

When the orbit is not in a bursting event, noise or parameter mismatch keeps 
$|\vect{x}_\bot(t)|$ non-zero. Thus, when the chaotic dynamics places the orbit near the fixed point at some time $t_0$, the linear approximation to the absolute separation between the  real and ideal (no noise or mismatch) trajectories 
along the unstable direction is 
\begin{equation}
\Delta(t) = 2\left| u_\bot (t) \right| = \left| u_M(t)-u_S(t)\right| \cong 2\left|u_\bot(t_0)\right|\exp\left[\eta(t-t_0)\right], \label{eq:Delta} 
\end{equation}
where $|u_\bot (t_0)|$ is determied by the small noise and/or mismatch.
In order to determine $B$, we neglect the components of the trajectories in the stable plane, $|\vect{x}_\bot (t)| \cong |u_\bot (t)|$, an approximation that is justified by the strong contraction of the linear flow along the stable plane --- the stable component shrinks exponentially fast. We also assume that saturation happens when one of the trajectories $u_M(t)$ or $u_S(t)$ reaches a fixed distance $u_\text{max}$ from the stable plane, at time $t_\text{sat}$, either on the positive or on the negative side of the plane. That is, either $|u_M(t_\text{sat})|$ or $|u_S(t_\text{sat})|$ is equal to $u_\text{max}$. With these approximations, the absolute values of $u_M(t)$, $u_S(t)$, $u_\parallelslant (t)$, and $u_\bot (t)$, all grow exponentially with the same exponent $\eta$, and the size of the burst that started at $t_0$ is $B = \Delta(t_\text{sat})$. 

To find an approximate value for $\Delta(t_\text{sat})$ there are two cases to consider: i) when $|u_\parallelslant (t_0)| > |u_\bot (t_0)|$, the trajectories for the master and slave systems remain relatively close to each other, so they both saturate together and ii) when  $|u_\parallelslant (t_0)| < |u_\bot (t_0)|$, the distance between the trajectories grows faster than their center and one of either the master or the slave trajectory saturates before the other one does. 
In case i), the trajectories are on the same side of the stable plane. Let us assume, without loss of generality, that the master trajectory saturates first, on the positive direction, hence,
\begin{equation}
B = \Delta (t_\text{sat}) = u_\text{max} -\left[ u_\parallelslant (t_0)-u_\bot (t_0)\right] \exp\left[ \eta (t_\text{sat}-t_0)\right]. \label{eq:B1}
\end{equation}
Using the solution for the master unstable component, Eq.\ \eqref{eq:u_M}, 
we can eliminate the time exponential in Eq.\ \eqref{eq:B1} 
\begin{equation}
\exp\left[ \eta (t_\text{sat}-t_0)\right]  = \frac{u_M(t_\text{sat})}{ u_\parallelslant (t_0)+u_\bot (t_0)} = \frac{u_\text{max} }{ u_\parallelslant (t_0)+u_\bot (t_0)},
\end{equation}
to write 
\begin{align}
B &= u_\text{max} - u_\text{max}\left( \frac{ u_\parallelslant (t_0)-u_\bot (t_0) }{ u_\parallelslant (t_0)+u_\bot (t_0)}\right), \\ 
B &= \frac{2u_\text{max} u_\bot (t_0) }{ u_\parallelslant (t_0)+u_\bot (t_0) } \label{eq:B_sameside} .
\end{align}
Similar analysis for other possibilities in case i) gives 
\begin{equation}
B = \frac{2u_\text{max} \left| u_\bot (t_0)\right| }{\left| u_\parallelslant (t_0)\right|+\left| u_\bot (t_0) \right|} .
\end{equation}
Notice that the maximum value of $B$ in case i) corresponds to $u_\text{max}$, when $|u_\bot (t_0)| \lesssim |u_\parallelslant (t_0)| $.

In case ii) the trajectories are always on opposite sides of the stable plane and  the distance between them has a larger absolute value than their center, hence it grows faster (with the same exponent, the growth rate is proportional to the initial size) and both trajectories saturate nearly at the same time, one of them at $u_\text{max}$ and the other at $-u_\text{max}$, so that their final separation is  $B = 2u_\text{max}$. If saturation happens close to the edge of the 3-D chaotic attractor, $2u_\text{max}$ is the maximum size possible for a burst.  Hence, the PDF for the burst sizes in case ii) is roughly $2u_\text{max}$, which can be interpreted as the narrow peak at the end of the distribution observed experimentally in Fig.\ 2. In principle, it is also possible that once one of the trajectories saturates it will quickly be reinjected, before the other one has had the time to reach saturation. In this situation the burst size is smaller than $2u_\text{max}$, but large
 r than $u_\text{max}$ for trajectories that remain on opposite sides of the stable plane. 

Using Eq.\ \eqref{eq:B_sameside} for the burst size, we find an approximate form for the distribution of bursts that are of the type described by case i) above ( $|u_\parallelslant (t_0)| > |u_\bot (t_0)| $). We write the probability density for a burst of size $B$ as 
\begin{equation}
p_B(B) = \iint  \delta\left( B - \frac{2u_\text{max} u_\bot (t_0) }{ u_\parallelslant (t_0)+u_\bot (t_0)}\right) p_{u_\parallelslant (t_0)}\left[u_\parallelslant (t_0)\right] p_{u_\bot (t_0)}\left[u_\bot (t_0) \right] d\!u_\parallelslant (t_0) d\!u_\bot (t_0), \label{eq:p_B1}
\end{equation}
where 
$\delta(\:\:)$ is a Dirac delta function, $p_{u_\parallelslant (t_0)}\left[u_\parallelslant (t_0)\right]$ and $p_{u_\bot (t_0)}\left[u_\bot (t_0) \right]$ are the probability densities, respectively, for the initial center and initial separation between the trajectories of the master and slave subsystems after reinjection. Equation \eqref{eq:p_B1} is equivalent to 
\begin{multline}
p_B(B) = \iint  \frac{2u_\text{max} u_\bot (t_0) }{B^2} \delta\left( u_\parallelslant (t_0)+u_\bot (t_0) - \frac{2u_\text{max} u_\bot (t_0) }{B}\right) p_{u_\parallelslant (t_0)}\left[u_\parallelslant (t_0)\right] p_{u_\bot (t_0)}\left[u_\bot (t_0) \right] \\  d\!u_\parallelslant (t_0) d\!u_\bot (t_0). \label{eq:p_B}
\end{multline}
Assuming that the distribution of the center of reinjection is a smooth and slowly varying function of the distance to the plane, and recalling that the initial separation between the trajectories $u_\bot (t_0)$ is small, 
\begin{equation}
p_{u_\parallelslant (t_0)} \left(\frac{2u_\text{max} u_\bot (t_0) }{B} -u_\bot (t_0) \right) \cong  p_{u_\parallelslant (t_0)}(0),
\end{equation}
in Eq.\ \eqref{eq:p_B}, such that 
\begin{align}
p_B(B)  &\cong \frac{2u_\text{max}  }{B^2} p_{u_\parallelslant (t_0)}(0) \int u_\bot (t_0)  p_{u_\bot (t_0)}\left[u_\bot (t_0) \right] d\!u_\bot (t_0),\\ 
  &= \frac{2u_\text{max}  }{B^2} p_{u_\parallelslant (t_0)}(0) \overline{u}_\bot (t_0), \label{eq:pB_powerlaw} \\
&= C B^{-2}, \label{eq:power_law}
\end{align}
where $\overline{u}_\bot (t_0)$ is the average value of the distance between reinjected trajectories, which is a constant, and $C =2u_\text{max} p_{u_\parallelslant (t_0)}(0) \overline{u}_\bot (t_0) $ is also constant. This last equation (Eq.\ \eqref{eq:power_law}) gives the power law with expoenent $-2$ observed experimentaly.

\section{Numerical Results}
\begin{figure}[!htbp]
\resizebox{12cm}{!}{\includegraphics{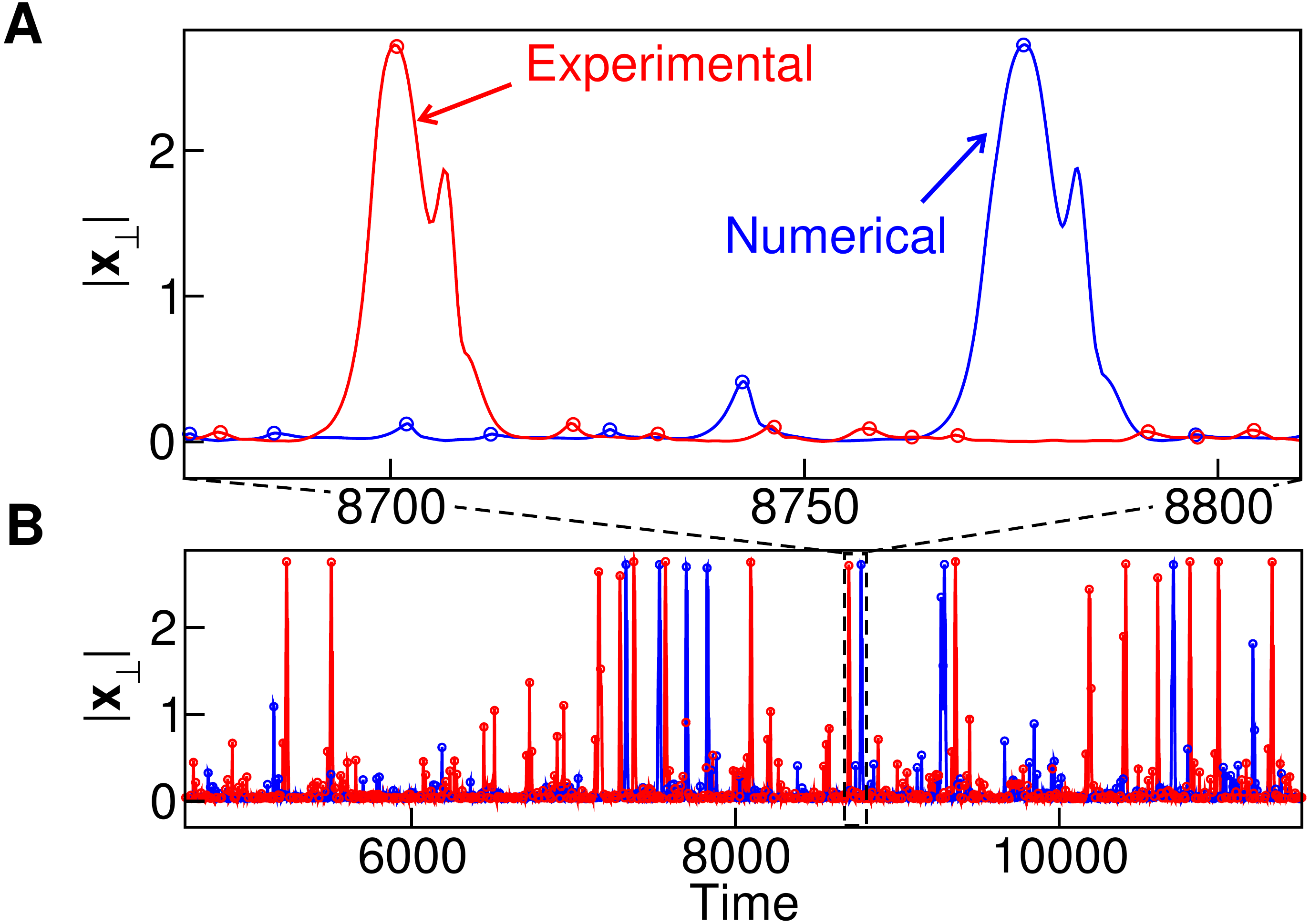}}
\caption{{\bf Comparison between experiment and the predictions of Eqs.\  (1--3)} {\bf A}, Temporal evolution of $\left| \vect{x}_\bot \right|$ showing the detail of a large bubble observed in the experiment (red) and resulting from the model (blue). Circles indicate the largest peaks $\left| \vect{x}_\bot \right|_n$, detected within each burst. {\bf B}, The same time series viewed over a longer time scale. The values of the model parameters fot the master oscillator are: $V_d = 0.58$ V, $R = 2,345\ \ \Omega$, $I_d = V_d/R = 0.25$ mA, $I_r = 22.5\ \ \mu$A, $\alpha_f = 11.60$, $\alpha_r = 11.57$, $R_1 = 1.298$, $R_2 = 3.44$, $R_3 = 0.043$, $R_\text{dc} = 0.150$, $R_4 = R_3+R_\text{dc} = 0.193$, and for the slave oscillator are: $V_d = 0.58$ V, $R = 2,345\ \ \Omega$, $I_d = V_d/R = 0.25$ mA, $I_r = 22.4\ \ \mu$A, $\alpha_f = 11.50$, $\alpha_r = 11.71$, $R_1 = 1.308$, $R_2 = 3.47$, $R_3 = 0.043$, $R_\text{dc} = 0.152$, $R_4 = R_3+R_\text{dc} = 0.195$, $c = 4.6$, $K_{ij} = 1$ for
  $i=j=2$ and 0 otherwise, and $c_\text{DK} = 0$. \label{fig:numerical_ts}}
\end{figure}

All the results shown in this article are experimental observations, for which
the model discussed in the previous section is in excellent agreement. Here we show a
comparison of the experiment and predictions of the model, obtained by numerical integration of
Eqs. (3-5). In order to induce bubbling, we modified the parameters of the slave circuit by 1\%.
Figure\ \ref{fig:numerical_ts} shows both numerical prediction and experimental observations of the temporal
evolution of $\left| \vect{x}_\bot \right|$. Over a short time scale, we observe that the shapes of the individual pulses of
desynchronization events are very similar in both theory and experiment, as shown in Fig.\ \ref{fig:numerical_ts}(a),
while Fig.\ \ref{fig:numerical_ts}(b) shows that, over a longer time scale, the statistics of these pulses in the
numerical simulation also appears to be similar to the experimental observation.

In Fig.\ \ref{fig:numerical_ts} colored circles show the positions and values of the largest peaks  $\left| \vect{x}_\bot \right|_n$ detected within each
desynchronization burst, which we define as the event-size. To simplify the algorithm to locate those points we
run a search for local maxima over a smoothed time series, where a maximum is defined as a point which is
higher than the 3 previous and the 3 consecutive points. Once a maximum higher than a certain limit (  $\left| \vect{x}_\bot \right| >1$) is located, any maximum appearing shortly after (7 time units) is rejected, to prevent single burst from
contributing with multiple event-sizes. This procedure is applied to both experimental and numerical data to
generate the sequence $\left| \vect{x}_\bot \right|_n$.

\begin{figure}[!htbp]
\resizebox{14cm}{!}{\includegraphics{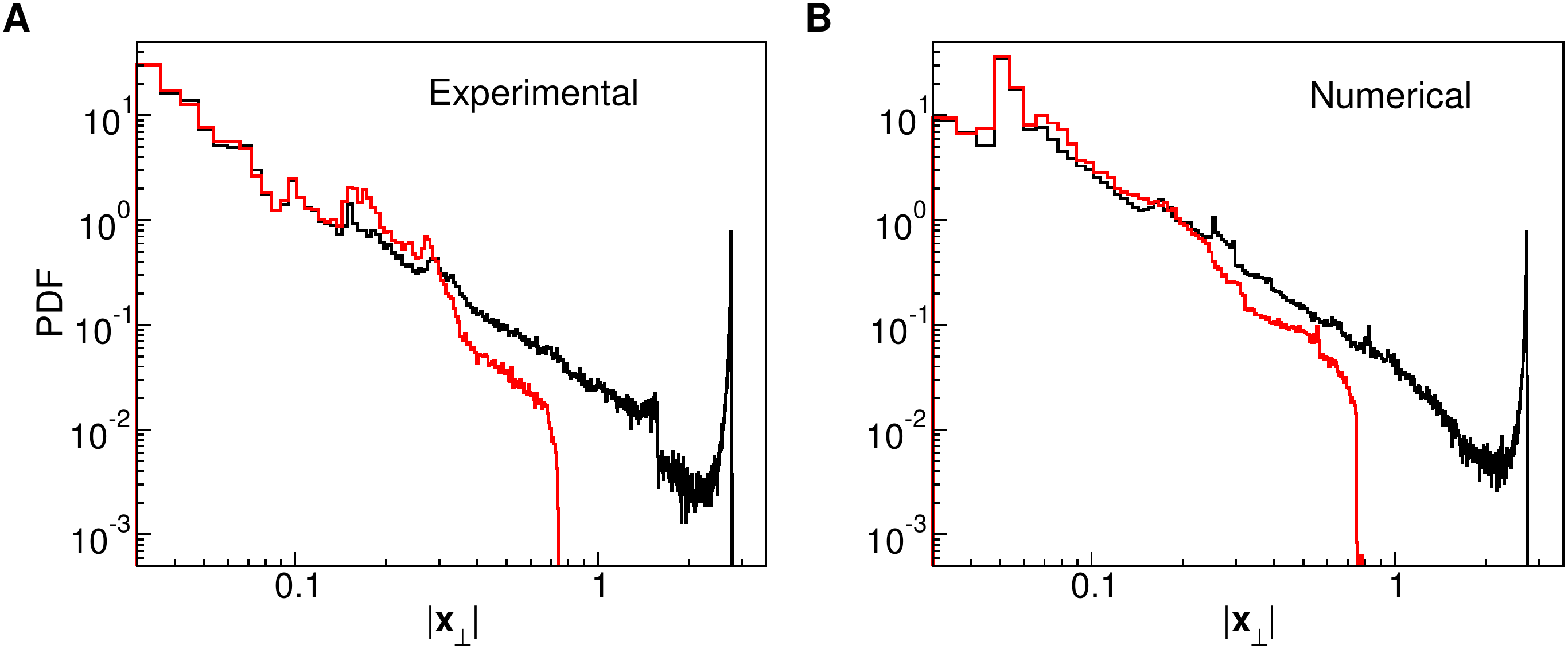}}
\caption{{\bf Comparison of the PDFs for the maxima of  $\left| \vect{x}_\bot \right|$} obtained ({\bf A}) in the experiment and ({bf B}) numerical integration of the model (Eqs.\ (1--3)). The values of the model parameters are the same given in the caption of Fig.\ \ref{fig:numerical_ts} except for $c_\text{DK}$, which is zero for the black curves and $c_\text{DK}=0.55$ for the red curves, $\left| \vect{x}_M \right|_\text{th} = 0.50$ in the experiment and $\left| \vect{x}_M \right|_\text{th} = 0.32$ in the model, and $ \left( K_\text{DK}\right)_{ij} = 1$ for $i=j=1$ and 0 otherwise. \label{fig:numerical_distribution}}
\end{figure}

With regards to the event-size distributions, there is great similarity between experimental observations
and theoretical predictions, as shown in Fig.\ \ref{fig:numerical_distribution}. In particular, the slope of the power law for intermediate-size events is $-2.0 \pm 0.1$ for both the experiment and for the theoretical model. Furthermore, both distributions show pronounced dragon-kings. When controlling dragon-kings, we find that precise value of the cut-off in the
distribution is quite sensitive to the value of $\left| \vect{x}_M \right|_\text{th}$ . We therefore take this as an adjustable parameter in the model and obtain close agreement between the observations and the model when we take the threshold at 0.32 for the model.

\section{Test of Statistical Convergence}
\begin{figure}[!htbp]
\resizebox{12cm}{!}{\includegraphics{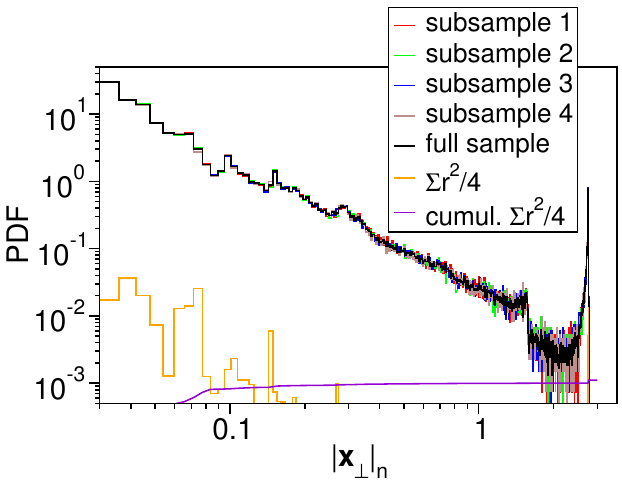}}
\caption{{\bf Probability Density Functions obtained with increasing sample size} converge to a stationary value.
Here the full sample has $N = 10^6$ points and produces the normalized histogram in black. The histograms constructed
with 4 subsamples comprised of $N/4$ points each deviate from the histogram of the full sample with a residual whose
average square value is plotted in the orange curve. The cumulative mean-square deviation is shown the violet curve,
ending in $\sim 10^{-3}$, which corresponds to a cumulative error in probability of 3\%. \label{fig:stat_conv}}
\end{figure}

The probability density function (PDF) of the experimental event sizes was
constructed by normalization of a histogram with a total of nearly one million events. These events will be referred here as the ``full sample'' and the number of events is the full sample size. We observed that histograms constructed with smaller subsamples, with one quarter of the total number, deviate only slightly from the histogram constructed from the full sample.
An approximate functional form for the PDF of event sizes was derived in Sec.\ \ref{sec:theory}. The hypothesis that we want to test here is that the values of the normalized histograms converge to some exact PDF which is not known from first principles with the increase of the number of points in the sample. This happens if each value of the histogram at a given bin is a random variable with finite average.
To assess this convergence quantitatively we calculate the square of the residuals, defined as the difference between the PDF obtained with one of the subsamples and the PDF of the full sample, average over the samples and take the square root of this
value, integrated over the whole domain of the PDFs, as an estimate of the total error in the probability calculated with one of the subsamples. The numerical value for this deviation in our case is 3\%. This confirms that the reported PDF is well-defined and characterizes the stationary statistical properties of the random variable $\left| \vect{x}_\bot \right|_n$. The normalized histograms constructed with 4 subsamples and the full sample that originated Fig.\ 2, and the square of the residuals and the cumulative square of the residuals are shown here in Fig.\ \ref{fig:stat_conv}.

\section{Statistical Test of Distribution}

\begin{figure}[!h!pbt]
\resizebox{14cm}{!}{\includegraphics{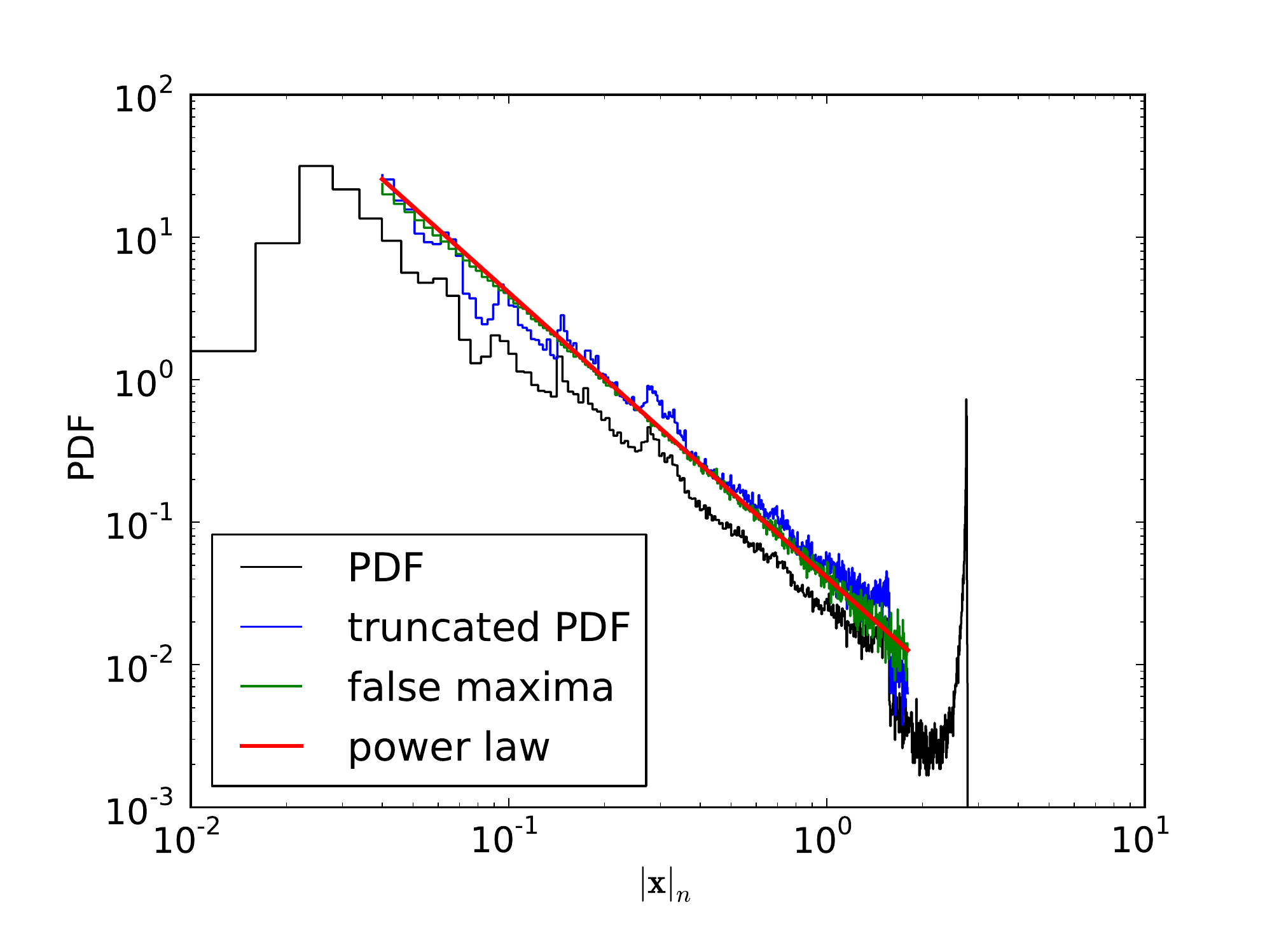}}
\caption{{\bf Probability Density Functions } for all the experimental event-sizes $|\vect{x}_\bot|_n$ (black); for events only in the range $0.04< |\vect{x}_\bot|_n<1.80$ (blue); the reference distribution (red), which is a truncated power law with exponent $-2$; and false maxima (green), randomly drawn according to the reference distribution.
The number of data points in the truncated distributions is $N \cong 3 \times 10^5$. \label{fig:truncated_power_law}}
\end{figure}

The theoretical analysis presented in Sec.\ \ref{sec:theory} suggests that the distribution of burst sizes in our system follows a power law with exponent $-2$, within a certain range of burst sizes, and ends in a statistically significant deviation upward from a power law, that we identify as due to the dragon-kings. There are a number of statistical tests in the literature that aim to verify whether data observed experimentally is consistent with a given statistical model (see supplemental Refs.\ [31, 32]). The idea behind all those tests is to check the hypothesis (called the \emph{null hypothesis}, in the language of statistics) that the empirical data was drawn from a reference distribution, prescribed by the model. We use the method known as Kolmogorov-Smirnov (K-S) test [31, 32], which is based on the maximum value of the absolute difference between the cumulative reference distribution and the cumulative empirical distribution (a normalized cumulative histogram constr
 ucted from the observed data). 
Figure 2 suggests the our data is nearly a power law for event sizes in the interval $0.04< |\vect{x}_\bot|_n<1.80$, Thus, in order to perform the K-S test, we first generate the normalized histogram of event sizes restricted to this interval. Our hypothesis is that, for the data restricted to this interval, the reference distribution is a truncated power law. 
Figure \ref{fig:truncated_power_law} shows the empirical distribution of the original, unrestricted data (black, the same curve shown in Fig.\ 2); the normalized histogram constructed using only the maxima that happen to fall within the range $0.04< |\vect{x}_\bot|_n<1.80$ (blue); and the reference distribution (red) which is a truncated power law. The integral of each of these PDFs is normalized to unit probability, hence all these curves show equal area in linear scale. 
We construct our empirical histograms using 500 bins with uniform sizes in a linear scale,  and we have no empty bins in the resulting histograms. It is usually suggested to use logarithmic bin sizes to avoid empty bins in heavy-tailed distributions, but this does not affect our case, because of the  large amount of bins and data points.
Using a least-squares fit to a straight line in log-log scale, we find the exponent of the truncated power law is always within the range $-1.89$ to $-2.14$, for different choices of the endpoints of the interval of $|\vect{x}_\bot|_n$, consistent with the theoretical value $-2$. We fix the endpoints at the values given and use the exponent $-2.0$ to define the reference distribution. 
To check that data drawn from the reference distribution passes the test, we produced synthetic data (false maxima) following the reference distribution and construct an ``empirical'' distribution from a sample of these false maxima with the same size $N$ as the true data. The PDF of false maxima is shown in green in Fig.\ \ref{fig:truncated_power_law}, where we see that it follows the reference distribution, but with large fluctuations that appear larger at the tail of the empirical distribution. 

\begin{figure}[!p!bht]
\resizebox{14cm}{!}{\includegraphics{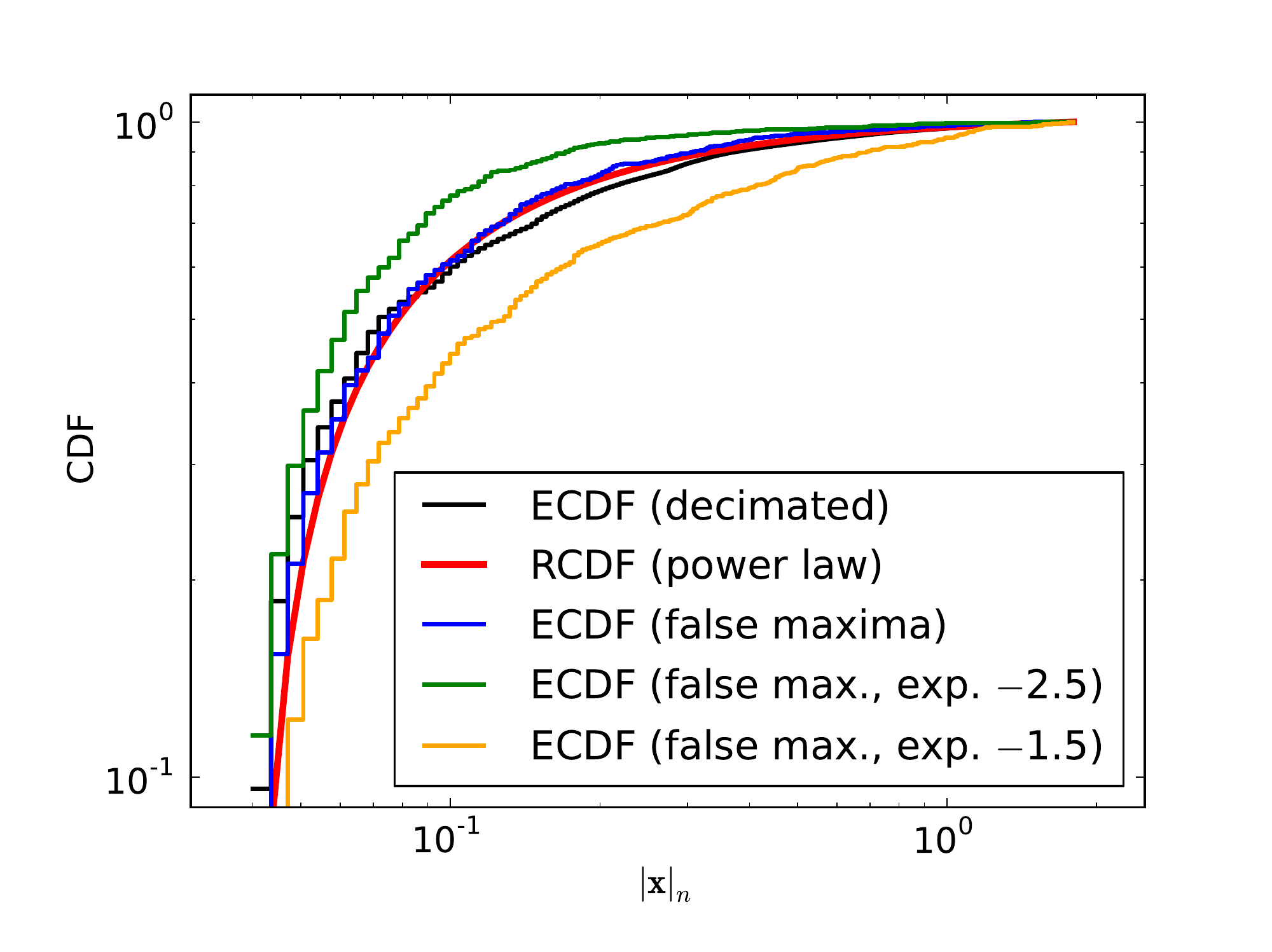}}
\caption{{\bf Cumulative Density Functions (CDFs)} for maxima in the truncated distribution limited to the interval $0.04< |\vect{x}_\bot|_n<1.80$ (black); for the reference distribution (red), which is a power law with exponent -2 truncated in this interval; for the artificially generated samples following the reference distribution (blue). 
The number of data points in the truncated distributions is $N \cong 10^5$. \label{fig:CDF}}
\end{figure}

The one-sample K-S test uses the maximum absolute value of the vertical distance $D$ between the empirical cumulative distributions functions (ECDF) and the reference cumulative distribution function (RCDF) as a statistic to decide whether the data was drawn from the reference distribution, thus, we proceed by integrating the empirical and reference PDFs up to size $|\vect{x}_\bot|_n$ to generate the corresponding empirical CDF (ECDF) and reference CDF (RCDF), shown in Fig.\ \ref{fig:CDF}. We see in Fig.\ \ref{fig:CDF} that the ECDF (blue) deviates significantly from the RCDF (red) in the range 
$|\vect{x}_\bot|_n \in \left[ 0.1 , 0.3 \right]$. 
The deviation is large enough that it enables one to reject the null hypothesis. The p-value for the distance $D$ observed in our ECDF ($\sqrt{N}D \sim 34$) is  negligible (0.0). In Fig.\ \ref{fig:CDF} we also see that the distribution for false maxima follows the reference distribution very accurately ($\sqrt{N}D \sim 0.72$) passing the test with p-value of 68\%. Synthetic data generated from power laws with different exponents $-1.5$ and $-2.5$ are also rejected, with even larger values of the statistics: $\sqrt{N}D \sim 124$ and $\sqrt{N}D \sim 92$, respectively. These results enable us to say that the bumps and valleys seen in the empirical distributions of the experimental data are not statistical fluctuations of a pure power law, but rather statistically significant structures in a more complicated distribution. The large peak at the end of the  distribution of unrestricted data causes an even larger deviation.

Notice that these results do not invalidate the analytical model presented in Sec.\ \ref{sec:theory}. The purpose of this model is to give a qualitative picture of the phenomenon of bubbling with a solvable distribution. The approximations and hypothesis made, although good for most of the events, do not take into account details of the nonlinear dynamics happening far from the unstable saddle point, which are determinant for the details of the distribution.

\begin{figure}[!bpt]
\resizebox{14cm}{!}{\includegraphics{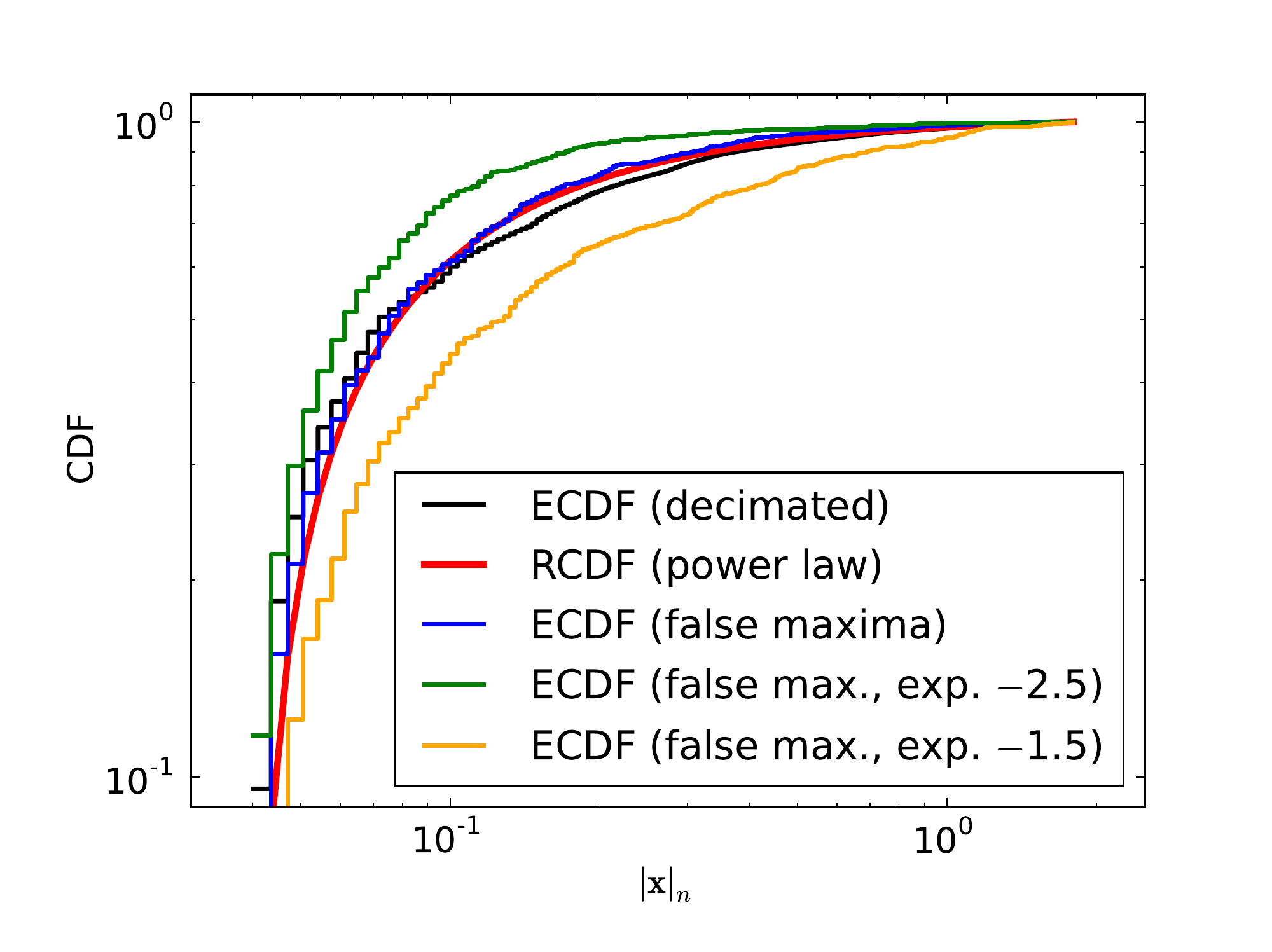}}
\caption{{\bf Cumulative Density Functions (CDFs)} for decimated maxima in the truncated distribution limited to the interval $0.04< |\vect{x}_\bot|_n<1.80$ (black); for the reference distribution (red), which is a power law with exponent -2 truncated in this interval; for the artificially generated samples following the reference distribution (blue); and for artificially generated samples following truncated power laws with exponent $-1.5$ (orange) and $-2.5$ (green).
The number of data points in the truncated distributions is $N \cong 10^3$. \label{fig:CDF_decimated}}
\end{figure}

In order to verify the consistency of our data with a power law we resample the raw data by decimating the sequence. The decimated sequence discards 700 maxima for each maximum kept. Thus we have only $N \sim 10^3$ in the decimated sample.  
The K-S statistic for this decimated sequence is $\sqrt{N}D \sim 1.3$, which gives a p-value of 8\%, thus passing the test. Synthetic data drawn from power laws with exponents $-1.5$, $-2.0$, and $-2.5$, and the same sample sizes as the decimated data give statistics$\sqrt{N}D \sim 5.3$, $\sqrt{N}D \sim 0.60$, and $\sqrt{N}D \sim 3.7$, respectively, which correspond to p-values $1.9 \times 10^{-24}$, 86\%, and $2.3 \times 10^{-12}$, respectively. It is clear that only the data drawn from the power law with exponent very close to $-2$ and our empirical data from the decimated sequence pass the test. The resulting cumulative distributions are shown in Fig.\ \ref{fig:CDF_decimated}, where we can see that the typical fluctuations of the empirical CDF obtained with the decimated data  have sizes similar to the deviations observed in CDF of the synthetic data, but the distributions with wrong exponent lie much above or below the reference distribution.

\section*{Additional References}
[31] A. Clauset, C. R. Shalizi, and M. E. J. Newman, SIAM Review {\bf 51}, 661 (2009). 

[32]  M. E. J. Newman, Contemporary Physics {\bf 46}, 323 (2005).
